\documentclass{aa}

\usepackage{graphics}
\usepackage{longtable}
\usepackage{txfonts}
\usepackage{graphicx}
\usepackage{hyperref}
\usepackage{multirow}
\usepackage{amsbsy}

\begin{document}

\title{Resolving the dusty circumstellar environment of the A[e] \\ supergiant HD 62623 with the VLTI/MIDI\thanks{Based on observations made with ESO Telescopes at Paranal Observatory under programs 078.D-O511 and 080.D.0181.}}

\authorrunning{Meilland et al.}
\titlerunning{VLTI/MIDI observations of the A[e] supergiant HD 62623}

\author{A. Meilland \inst{1}, S. Kanaan\inst{2}, M. Borges Fernandes\inst{2}, O. Chesneau\inst{2}, F. Millour  \inst{1},  Ph. Stee \inst{2}, and B. Lopez \inst{2}}

   \offprints{meilland@mpifr-bonn.mpg.de}

\institute{
Max Planck Intitut f\"ur Radioastronomie, Auf dem Hugel 69, 53121 Bonn, Germany 
\and
UMR 6525 CNRS H. FIZEAU Ð UNS, OCA, Campus Valrose, F-06108 Nice cedex 2, France,  CNRS - Avenue Copernic, Grasse, France.}

   \date{Received; accepted }

   \abstract{B[e] stars are hot stars surrounded by circumstellar gas and dust responsible for the presence of emission lines and IR-excess in their spectra. How dust can be formed in this highly illuminated and diluted environment remains an open issue.}
  {HD~62623 is one of the very few A-type supergiants showing the B[e] phenomenon. We studied the geometry of its circumstellar envelope in the mid-infrared using long-baseline interferometry, which is the only observing technique able to spatially resolve objects smaller than a few tens of milliarcseconds.}
{We obtained nine calibrated visibility measurements between October 2006 and January 2008 using the VLTI/MIDI instrument in SCI-PHOT mode and PRISM spectral dispersion mode with projected baselines ranging from 13 to 71~m  and with various position angles (PA). We used geometrical models and physical modeling with a radiative transfer code to analyze these data.}
{The dusty circumstellar environment of HD~62623 is partially resolved by the VLTI/MIDI even with the shortest baselines. The environment is flattened (a/b$\sim$1.3$\pm$0.1) and can be separated into two components: a compact one whose extension grows from 17~mas at 8$\mu$m to 30~mas at 9.6$\mu$m and stays almost constant up to 13$\mu$m, and a more extended one that is over-resolved even with the shortest baselines. Using the radiative transfer code MC3D, we managed to model HD~62623's circumstellar environment as a dusty disk with an inner radius of 3.85$\pm$0.6 AU, an inclination angle of 60$\pm$10$^o$, and a mass of 2.10$^{-7}$M$_\odot$.}
{It is the first time that the dusty disk inner rim of a supergiant star exhibiting the B[e] phenomenon is significantly constrained. The inner gaseous envelope likely contributes up to 20$\%$ to the total N band flux and acts like a reprocessing disk. 
Finally, the hypothesis of a stellar wind deceleration by the companion's gravitational effects remains the most probable case since the bi-stability mechanism does not seem to be efficient for this star.}


   \maketitle
%

\section{Introduction}

\begin{figure*}[t]
\centering   \includegraphics[height=6cm]{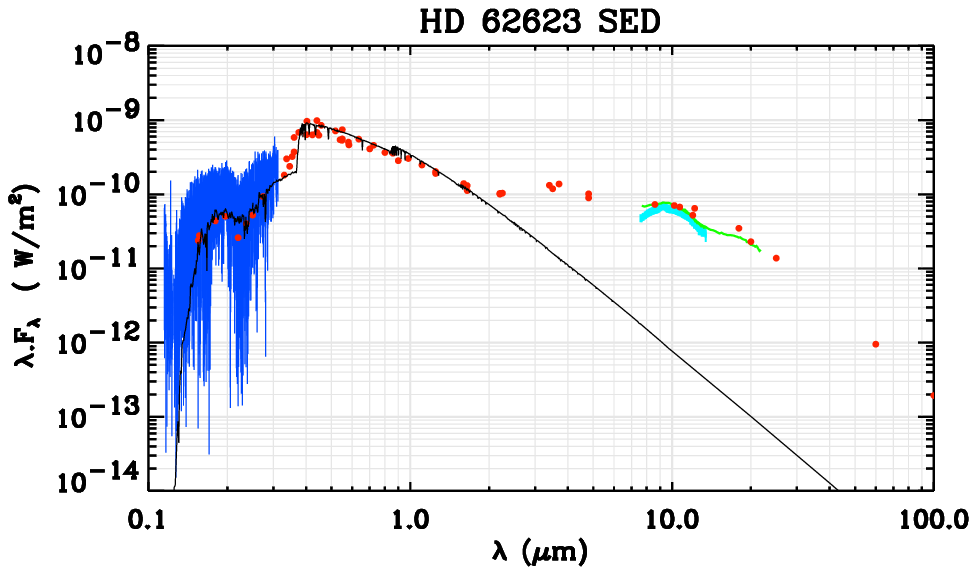}
\includegraphics[height=6cm]{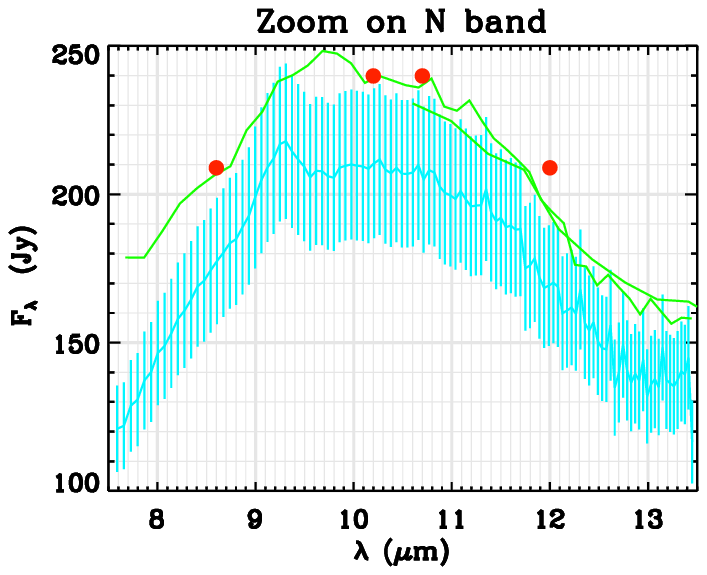}
\caption{Left : HD~62623 SED from various sources of literature: red diamonds are flux measurements listed in Bittar et al. (2001); green lines, IRAS LRS spectra; dark blue lines, IUE spectra; light blue lines with errors bars, our VLTI/MIDI measurements; and the dark line is the best reddened Kurucz model, whose parameters are presented in Table~1. Right : Zoom of the N band of the SED (8-13$\mu$m) to exhibit differences between the various measurements obtained in this spectral band and the VLTI/MIDI spectra.}
\label{SED}
\end{figure*}

The B[e] phenomenon is defined by strong Balmer lines in emission, low excitation emission lines (especially of ionized metals such as FeII), forbidden emission lines of [FeII] and [OI], and strong infrared excess due to hot circumstellar dust (Lamers et al. 1998). However, these objects are not a homogeneous group of stars in terms of stellar evolution, as such features have been detected for both pre-main sequence and evolved stars. Thus, this group of stars has been divided into four sub-classes : B[e] supergiants, Herbig AeB[e] stars, compact planetary nebulae B[e]-type stars, and symbiotic B[e]-type stars. Moreover, since the spectral class of some of these stars was still very difficult to determine accurately, a fifth class was added: unclassified B[e] stars. Recently,  in an effort to investigate this mysterious sub-group, Miroshnichenko (2007) proposed to classify most of their members as a new homogeneous group of non-supergiant and non-pre-main sequence stars. He proposed to name this group after its precursor, i.e. FS CMa's group. Finally, the binarity detected for 30\% of these stars could play a key role for the origin of the B[e] phenomenon for these FS CMa stars.

In this context, HD~62623 (3 Puppis, HR 2996) was classified as an A2Iabe in the \textit{Yale Bright Stars Catalogue} (Hoffeit \& Jaschek 1982) making it one of the latest stars showing the B[e] phenomenon. Other classifications range from A2Ia to A3IIpe. Its distance was estimated to be 700 $\pm$ 150 pc, and its radius R$_\star$=55$\pm$10R$_\odot$ (Bittar et al. 2001). It exhibits radial velocity variations with a period of 137.7 days, interpreted as direct evidence for an unseen companion (Lambert 1988). Plets et al. (1995) found that a 161.1-day period can also fit these variations. In both hypotheses, the system seems to be composed of a massive primary star with 31~M$_\odot<~$M$_\star~<~$39~M$_\odot$ and a small mass ratio of 0.03~$<~$M$_2/$M$_1~<~$0.15, while the projected semi-major axis is 1.6~AU~$<$~a.sin~i~$<$~2.4 AU.  

Using color indices (H - K)  and (K - L), Allen (1973) found that HD~62623's infrared excess is too large to only originate from the free-free and free-bound emissions of a gaseous envelope and is probably due to the presence of optically thick dust in the circumstellar environment of this star. Observations with the Low-Resolution Spectrograph (LRS) of the InfraRed Astronomical Satellite (IRAS) clearly exhibit a 9.6 $\mu$m silicate feature in the object spectra, confirming the presence of dust in the envelope. Since the star also shows evidence of H$\alpha$ and [OI] emission lines (Swings \& Allen 1976), it was classified as a B[e] supergiant, i.e. sgB[e] (Miroshnichenko 2007).

The formation mechanisms responsible for sgB[e] stars' circumstellar environment are still an open issue. Spectroscopic and polarimetric observations suggest that their circumstellar envelopes are non-spherical (Magalh\~aes 1992). Zickgraf et al. (1985) proposed a general scheme for these stars consisting of a hot and fast line-driven polar wind responsible for the presence of forbidden lines and a slowly expanding dense equatorial region where permitted-emission lines and dust can form. The deviation from the spherical geometry for this model can be reproduced by a nearly critically rotating star. 

Using aperture masking interferometry with the Keck I telescope, Bittar et al. (2001) partially resolved HD~62623's circumstellar envelope between 1.65 and 3.08 $\mu$m and showed that the envelope extension was increasing from 6-7 mas at 1.65 $\mu$m to 25-30 mas at 3.08 $\mu$m. Stee et al. (2004) have successfully modeled both these observations and the Spectral Energy Distribution (SED) using a gas + dust model with an inner dust radius of 15-20R$_\star$ and a temperature at this boundary of 1500K. In their best model, the gas emission originates from the inner part of the envelope and not from a polar wind. Nevertheless, the issue of HD~62623's envelope flattening remains unsolved, even if Yudin \& Evans (1998) measured a polarization in the visible of 1.5\% with a polarization angle of 95$\pm$5$^o$, which seems to favor a non-spherical object.

By putting strong constraints on the envelope geometry, long baseline infrared interferometry can be a key technique to study the physics of active hot stars (Stee et al. 2005). Using both VLTI/MIDI and VLTI/AMBER data, Domiciano de Souza et al. (2007) resolved the circumstellar environment of the sgB[e] CPD -57$^o$ 2874 and modeled it as a flattened compact envelope with an extension increasing with wavelength.

In this paper, we present new spectrally resolved VLTI/MIDI interferometric observations of HD~62623 that allow us to characterize its circumstellar envelope geometry and determine its physical parameters. 

The paper is organized as follows. In Section 2 we start our modeling of HD~62623 by reconstructing its SED in order to constrain its stellar parameters and distance. The VLTI/MIDI observations and the data reduction process are briefly introduced in Section 3, and a first analysis of the calibrated visibilities using simple geometrical models is presented in Sect. 4. The result of these study is used as a starting point for a three-dimensional Monte-Carlo radiative transfer modeling presented in Sect. 5. Finally, the physical properties of HD~62623 are discussed in Sect. 6, while Sect. 7 draws the main conclusions of this study.

\section{Preliminary study}

\subsection{The Spectral Energy Distribution}
In order to accurately determine the stellar parameters of HD~62623, we started our study by reconstructing the object SED using photometric and low resolution spectral measurements ranging from ultra-violet to far-infrared. We used data from various sources in the literature already described in Bittar et al. (2001). These data include: IRAS fluxes and Low Resolution Spectra (LRS), IUE spectra, and other photometric measurements in the UV, the visible, and the near-IR. We also included our N band VLTI/MIDI spectra in this study. The resulting reconstructed SED is plotted in Fig.~\ref{SED}. The strong IR-excess due to circumstellar dust is evidenced by the inflection point in the SED around 1.5 $\mu$m and by the 9.6 $\mu$m silicate emission band visible on IRAS LRS and VLTI/MIDI spectra between 8 and 13 $\mu$m. The VLTI/MIDI flux is of the order of the other mid-IR measurements. This clearly shows that most of the N-band emission originates from the 250~mas interferometric field of view of this instrument. We note that a second silicate emission band is also visible in IRAS LRS spectra around 20 $\mu$m. 

\subsection{Circumstellar or interstellar extinction}

We have tried to fit the UV and visible parts of the reconstructed SED using typical Kurucz models (Kurucz 1979) for supergiant stars; i.e, with 1.5 $<$ log~g $<$ 2.5 and 8000K $<$ T$_{eff}$ $<$ 10000K. None of these models were able to simultaneously fit all the SED from UV wavelengths up to 1 $\mu$m, and the observed SED seems to be slightly reddened. To explain this reddening, two hypotheses can be formulated: (1) if the circumstellar dust becomes optically thick at short wavelengths, it can absorb a part of the stellar UV emission, assuming this dust is located in the line of sight, or (2) the reddening can be due to interstellar matter.

It is hard to discriminate between these two hypotheses without fully modeling HD~62623's circumstellar environment. However, it is still possible to estimate the reddening without solving the issue concerning its origin. Using Meynet \& Hauck (1985) intrinsic Geneva colors laws for A-F supergiants, Plets et al. (1995) estimated a reddening E(B-V) of 0.17$\pm$0.03.

\begin{table}[!t]
{\centering \begin{tabular}{cc}
\hline Parameter & Value\\
\hline
T$_{eff}$ & 8250 $\pm$ 250 K\\
log~g & 2.0 $\pm$ 0.5\\
R$_\star$ & 65 $\pm$ 5 R$_\odot$\\
\textit{d}& 650 $\pm$ 100 pc\\
E(B-V) & 0.17 $\pm$ 0.03$^{(1)}$\\
\hline
\end{tabular}\par}
\label{stellar}
\caption{Stellar parameters, distance, and extinction determined from the fit of HD~62623 SED. (1) value from Plets et al. (1995)}
\end{table}

\subsection{Stellar parameters}

Assuming that the extinction originates from a standard interstellar medium (R$_V$=3.1) and that it follows Cardelli et al.'s (1989) law from UV to far-IR wavelengths, the SED can be successfully fitted with reddened Kurucz models with T$_{eff}$ ranging from 8000K to 8500K and log~g from 1.5 to 2.5. Our best model, using T$_{eff}$=8250K and log~g=2.0, is overplotted in Fig.~\ref{SED}. The good agreement between the observed SED and the modeled one is obtained for an \textit{d}/R$_\star$ ratio of 10$\pm$1, where the distance \textit{d} is expressed in parsecs and the stellar radius R$_\star$ in R$_\odot$. Since a typical value of R$_\star$ for a A supergiant is 65$\pm$5 (Allen 2000), we can estimate a distance for the star of 650$\pm$100~pc, which is compatible with the previous estimation by Bittar et al. (2001). The resulting parameters used in this paper to model HD~62623's central star are presented in Table 1.

\section{VLTI/MIDI Observations and data reduction}

\begin{table*}[t]
{\centering \begin{tabular}{ccccccc}
\hline Name & Obs. Date & Obs. Time & Stations & \multicolumn{2}{c}{ Projected  baseline} & Calibrators\\
&&(UT)&&Length(m)& P.A.($^o$)&\\
\hline
B$_1$&2006/10/17&06:42&D0-H0&49.5&27.2&$\gamma$ Eri, Procyon, $\beta$ Gru\\
B$_2$&2006/10/17&08:10&D0-H0&56.8&48.0&$\gamma$ Eri, Procyon, $\beta$ Gru\\
B$_3$&2006/12/16&04:00&G0-H0&27.8&45.8&Procyon, 4 Eri\\
B$_4$&2006/12/19&08:38&G0-K0&56.6&84.8&Alphard, Procyon\\
B$_5$&2006/12/21&08:19&G0-H0&29.0&84.0&$\gamma$ Eri, Procyon, Alphard\\
B$_6$&2007/10/04&08:22&E0-G0&13.3&39.4&Sirius\\
B$_7$&2007/12/10&05:22&G1-H0&71.4&-4.1&Sirius\\
B$_8$&2007/12/14&08:13&G1-H0&70.9&15.4&Sirius\\
B$_9$&2008/01/12&07:08&E0-G0&14.0&85.3&Sirius, Alphard\\
\hline
\end{tabular}\par}
\label{MIDI_log}
\caption{VLTI/MIDI Observations log of HD~62623 and its calibrators.}
\end{table*}

VLTI/MIDI observations of HD~62623 were carried out at Paranal Observatory between October 2006 and January 2008 (Guaranteed Time Observing runs 078.D-0511 and 080.D.0181) with the 1.8m Auxiliary Telescopes (ATs). We obtained nine visibility measurements with projected baselines ranging from 13.4 to 71.4 meters and with various orientations on the sky plane. Three stars were used as calibrators during this observing campaign: Procyon, Sirius, and Alphard. The log of these observations is presented in Table 2.

All observations were made using the SCI-PHOT mode, which enables a better visibility calibration since the photometry and the interferometric fringes are recorded simultaneously. Thanks to the PRISM low spectral dispersion mode, we also obtained spectrally resolved visibility with R=30 in the N band (7.5-13.5$\mu$m). Two different reduction packages were used to reduce these data : MIA developed at the Max-Planck Institut f\"ur Astronomie and EWS developed at the Leiden Observatory (MIA + EWS, ver.1.5.1). Since the two methods provide the same results within the error bars, we decided to use the MIA reduced visibilities in the following work.

\begin{figure}[!t]
\centering   \includegraphics[width=0.4\textwidth]{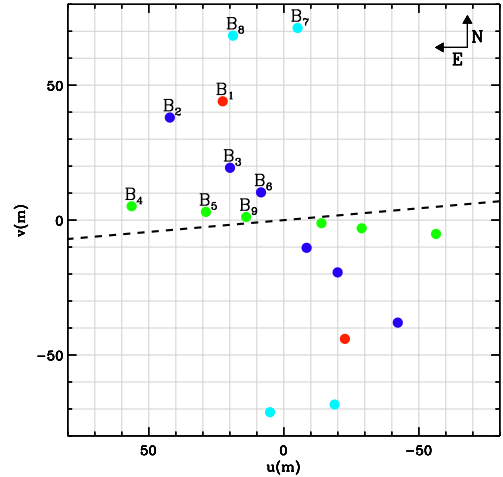}
\caption{UV plane coverage obtained  for HD62623. The dotted line represents the polarization angle measured by Yudin \& Evans (1998). The circle colors represent the two triplet of aligned baselines discussed in Sect. 3 : red for PA$\sim$40$^o$ and green for PA$\sim$85$^o$. }
\label{visi_uv}
\end{figure}

HD~62623's (u,v) plane coverage is plotted in Fig~\ref{visi_uv}. The large spread of baseline length (between 13.4 and 71.4) and P.A. (between -4.1$^o$ and 85.3$^o$)  will allow us to put strong constraints on the envelope geometry in the N band (i.e. extension, flattening, radial intensity profile). Moreover, two triplets of baselines are almost aligned:  B$_2$, B$_3$, and B$_6$ with P.A.$\sim$40$^o$, and B$_4$, B$_5$, and B$_9$ with P.A.$\sim$85$^o$. This will simplify our modeling by allowing us to directly determine the object projected intensity profile at these two orientations.

\begin{figure}[!t]
\centering   \includegraphics[width=0.46\textwidth]{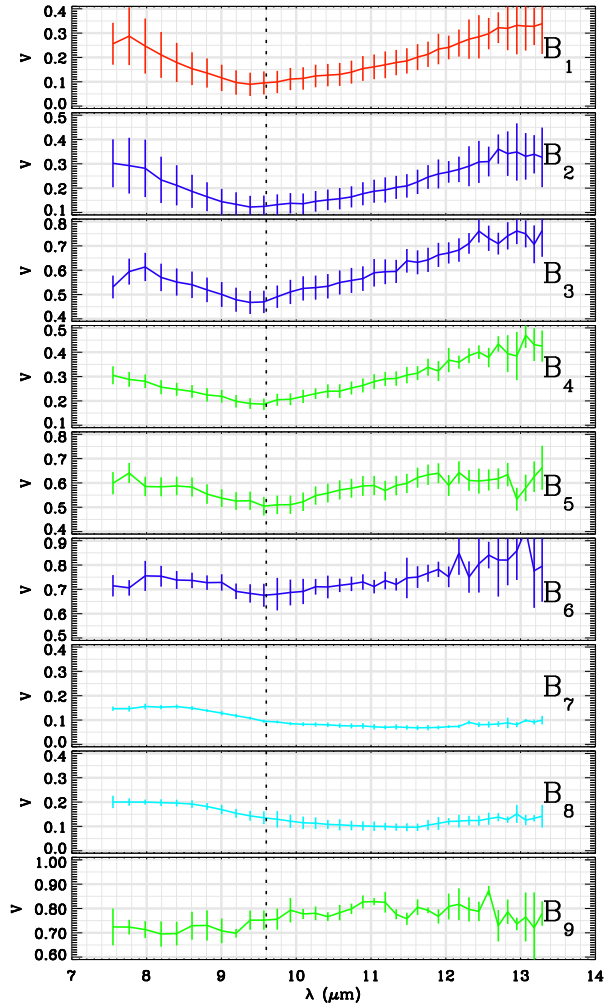}
\caption{HD~62623 VLTI/MIDI calibrated visibilities for the nine baselines from Fig. 2 plotted as a function of wavelength. The dotted line corresponds to the central wavelength of the silicate band, i.e, 9.6 $\mu$m.}
\label{visi}
\end{figure}

\section{Results}

\subsection{Visibilty and Gaussian equivalent FWHM}

The calibrated visibilities for the nine projected baselines are plotted as a function of wavelength in Fig.~\ref{visi}. All these spectrally resolved visibilities, except the ones from the two longest baselines (i.e., B$_7$ and B$_8$), exhibit a small drop with a minimum around 9.6 $\pm$0.1 $\mu$m. Such an effect, often found in N-band interferometric observations of dusty circumstellar environments, is due to an opacity increase between the continuum and a silicate band. This greater opacity in the silicate band causes the circumstellar envelope to appear more extended than in the continuum. The fact that such an effect is not visible in the B$_7$ and B$_8$ data probably means that both the silicate and the continuum emission are over-resolved with these baselines. We note that, for this A[e] supergiant star, the silicate band is clearly in emission (see Fig.~\ref{SED}). 

The object is partially resolved for all baselines, even for the shortest ones (i.e., V$<$0.9 for B$_6$=13.3m and B$_9$=14.0 m). This implies that the circumstellar envelope is quite extended. To roughly estimate its size, we calculate the Gaussian equivalent FWHM for each baseline and at three wavelengths (i.e., 8, 10, and 12 $\mu$m). The results, presented in Table~3, clearly show that the envelope extension is growing by about 45$\%$ between 8 and 10 $\mu$m, whereas it remains nearly constant between 10 and 12 $\mu$m. As we already mentioned, the envelope size variation between 8 and 10 $\mu$m is mainly due to the change of opacity between the continuum and the silicate band, but another effect has to be taken into account to explain why the size does not drop back to its 8 $\mu$m value from 10 to 12 $\mu$m. Remembering that larger wavelengths can probe cooler region it implies, for an untruncated gaseous or dusty envelope heated by a central source, i.e., a single star, that the size of the emission is growing with the wavelength as the cooler regions are located further from the heating source. If the size in the continuum were not increasing with the wavelength this would have implied that the disk was somehow truncated, as it was discovered for the classical Be star $\alpha$ Arae (Meilland et al. 2007), or that the disk was mainly isothermal (Jones et al. 2004).

\begin{table}
\renewcommand{\multirowsetup}{\centering}
{\centering \begin{tabular}{ccccc}
\hline  
&\multicolumn{3}{c}{Gaussian Disk FWHM (mas)} & P.A.($^o$)\\
				& 8$\mu$m& 10$\mu$m& 12$\mu$m & \\
\hline				
B$_4$&17.6$\pm$ 0.3&24.1$\pm$ 0.3&24.4$\pm$ 1.2&\multirow{3}{1.5cm}{P.A.$\sim$85$^o$}\\
B$_5$&21.5$\pm$ 0.7&30.0$\pm$ 1.5&32.3$\pm$ 1.0&\\
B$_9$&31.7$\pm$ 4.0&40.7$\pm$ 4.9&43.4$\pm$ 6.3&\\
\hline
B$_2$&16.6$\pm$ 2.6&26.4$\pm$ 1.0&26.1$\pm$ 1.5&\multirow{3}{1.5cm}{P.A.$\sim$40$^o$}\\
B$_3$&22.7$\pm$ 0.7&32.8$\pm$ 1.7&31.6$\pm$ 3.5&\\
B$_6$&43.8$\pm$ 3.8&58.7$\pm$ 4.8&59.2$\pm$ 5.9&\\
\hline
B$_1$&21.2$\pm$ 2.6&32.4$\pm$ 1.1&32.4$\pm$ 1.6&P.A.=27.2$^o$\\
B$_7$&16.8$\pm$ 0.7&23.8$\pm$ 0.7&27.7$\pm$ 1.0&P.A.=-4.1$^o$\\
B$_8$&15.4$\pm$ 0.6&22.4$\pm$ 0.6&24.9$\pm$ 0.7&P.A.=15.4$^o$\\
\hline
\\
\end{tabular}\par}
\label{gauss_estimation}
\caption{Gaussian disk equivalent FWHM for all baselines and at three different wavelengths. The baselines are classified by P.A. in order to exhibit the inconsistency of a single Gaussian modeling, which would have provided similar sizes for baselines with the same P.A.}
\end{table}

\begin{figure}[!t]
\centering   \includegraphics[width=0.45\textwidth]{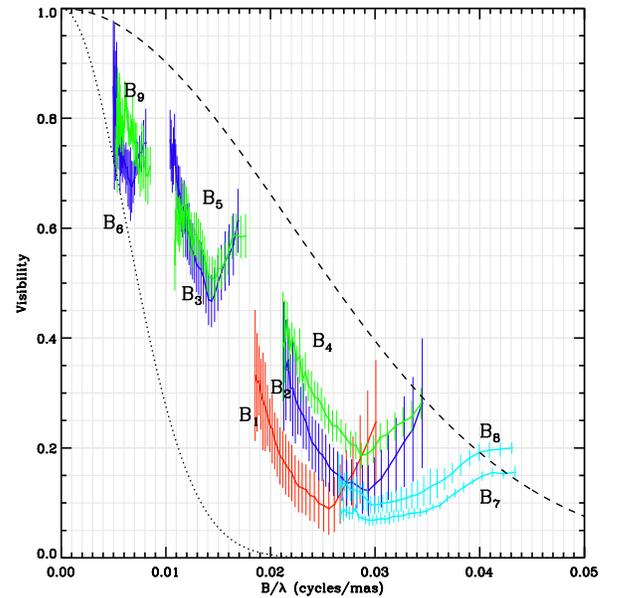}
\caption{HD 62623 VLTI/MIDI calibrated visibilities plotted as a function of
the spatial frequency B/$\lambda$. The colors correspond to the orientation of the projected baselines: green for 85$^o$,  blue for 40$^o$, and red for 27.2$^o$. 17~mas (dashed line) and 60~mas (dotted line) Gaussian disks are overplotted to illustrate the inconsistency of modeling HD~62623's circumstellar environment as a single Gaussian disk or ellipse.}
\label{all_visi_spafreq}
\end{figure}

\begin{figure*}[!t]
\centering   \includegraphics[width=0.99\textwidth]{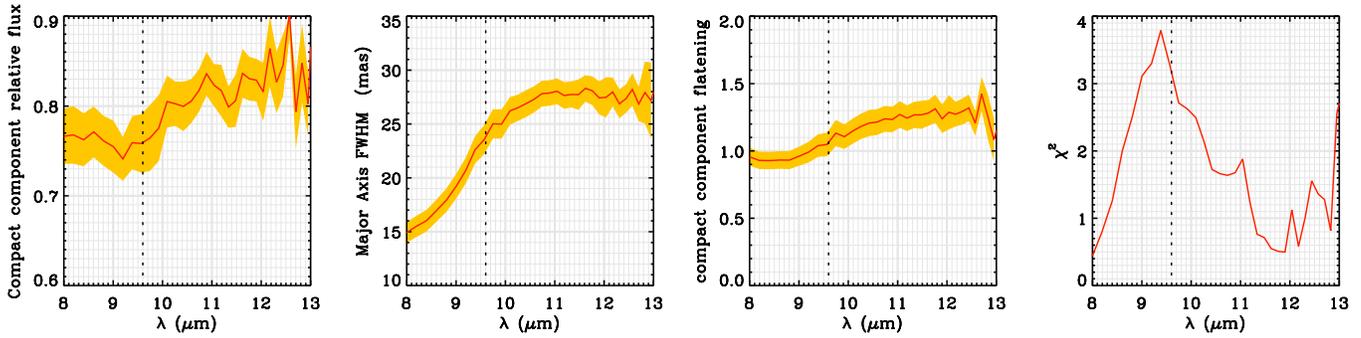}
\caption{Best-fit models parameters and $\chi^2$ for a fully resolved component + elliptical Gaussian geometry (G1). From left to right: Normalized flux of the most compact component F$_1$, FWHM of the major and minor axis of the compact component, and corresponding reduced $\chi^2$. The vertical dotted line indicates the position of the center of the silicate band, i.e 9.6 $\mu$m.}
\label{models}
\end{figure*}

If the intensity distribution of the object was close to an elliptical Gaussian, the estimated FWMH should only depend on the baseline orientation and not on its length. As we already mentioned in the previous section, two groups of  three baselines are almost aligned. Consequently, the Gaussian equivalent FWHM should be nearly constant within each one. As is shown in Table~3, this is not the case for HD~62623 and for a given wavelength and P.A. The estimated size shrinks with the baseline length and finally, the Gaussian equivalent size measured for the shortest baseline is about twice the one measured for the longest baseline. This result is also valid if we model the object as a single uniform ellipse or elliptical ring. This clearly shows that the object geometry is more complicated and should be modeled using at least two components: a compact one which is resolved only by the longest baselines and a more extended one which may even be fully resolved by the smallest baselines.

This assumption becomes obvious when plotting the visibility for all baselines as a function of the spatial frequency $|\overrightarrow{B}|/\lambda$ (Fig~\ref{all_visi_spafreq}). This plot also exhibits all the previously described characteristics of the object in a very synthetic way. The wavelength dependence is evidenced in each of the seven curves representing the visibility for each baseline. These curves are inverted compared to Fig~\ref{visi} (i.e., the wavelength increases from right to left), but both the silicate band and the continuum effects on the visibility are shown. Clues of the flattening of the envelope, at least in the silicate band, are also evidenced since the visibility depends not only on the baseline length but also on its orientation. The object is larger in the B$_2$, B$_3$, and B$_9$ orientation (P.A.$\sim$40$^o$) than in the B$_4$, B$_5$, and B$_9$ one (P.A.$\sim$85$^o$), and is even larger along B$_1$  (P.A.=27.2$^o$). These visibility variations as a function of the P.A. are consistent with an orientation of the major-axis of 5$^o$ derived from Yudin \& Evans (1998) polarization measurements. Nevertheless, the flattening cannot be accurately determined without knowing the radial intensity profile of the object. We note that B$_7$ and B$_8$ are almost aligned with the putative envelope major-axis.

\subsection{A two-component model}

It is obvious from Table 3 and Fig~\ref{all_visi_spafreq} that HD~62623 MIDI visibilities cannot be fitted by a single component model. Thus, we try to fit them with three different kinds of two-component geometrical models. For all of these models, the most compact component is an elliptical Gaussian defined by three parameters: the major-axis FWHM (a$_1$), the minor-axis FWHM (b$_1$), and the orientation of the major-axis into the sky plane ($\theta_1$). Since the polarization angle measurement seems to provide the orientation of the minor-axis with a good accuracy, we can set the value of $\theta_1$ to 5$^o$. We will see in Sect. 5 that this component probably corresponds to the inner rim of the circumstellar dusty disk and that it can be modeled as an elliptical ring. However, the envelope is not fully resolved and this component can be successfully fitted by several other flattened models including Gaussian or uniform ellipse. 

The last parameter concerning this compact component is the flux ratio between this Gaussian (F$_1$) and the total normalized flux at each wavelength. The flux of the extended component is consequently (1-F$_1$). Since we want to determine if it can be fully resolved by the smallest baselines, we model it following three different ways of increasing complexity to test if we can put some constraints on its geometry. To check the consistency of our two-component modeling, we also add a "reference" model assuming only one elliptical Gaussian. The parameters of the four different geometries tested (G0, G1, G2, and G3) are the following~:
\begin{itemize}
\item G0 : This is the ``reference" model. The geometry is defined by a single elliptical Gaussian. This model has only 2 free parameters : a$_1$, and b$_1$.
\item G1 : Here, we consider that the extended component is fully resolved, and we model it as a homogeneous constant contribution. This first model has 3 free parameters : a$_1$, b$_1$, and F$_1$.
\item G2 : We try to model the extended component as a Gaussian with the same flattening as the compact one so that it is defined only by its major axis FWHM a$_2$. This model has 4 free parameters  : a$_1$, a$_2$, b$_1$, and F$_1$.
\item G3 : In this last model, the minor-axis FWHM (b$_2$) of the extended component is also a free parameter. Thus, this model has 5 free parameters : a$_1$, a$_2$, b$_1$, b$_2$ and F$_1$.
\end{itemize}

Since the envelope extension varies within the N band, we assume that the free-parameters are all wavelength dependent. For each wavelength, we try to minimize the model reduced $\chi^2$ using a Levenberg-Marquardt 2D algorithm with 10$^3$ random starting positions. We finally calculate the global $\chi^2$ for all wavelengths to compare the best-fitted models of the three geometries (and the reference G0). The value of the global $\chi^2$ are 9.14 for G0, 2.62 for G1, 2.62 for G2, and 3.26 for G3. Thus, this modeling clearly shows the lack of data to constrain the geometry of the large component of our model but the consistency of a two-component modeling approach. More proof of the difficulty to constrain the large component geometry is the fact that the wavelength dependence of a$_2$ and b$_2$ for the best G2 and G3 models is erratic, with values randomly distributed between 100~mas and 1000~mas. Finally, it is obvious that we need more measurements with smaller spatial frequencies to constrain the geometry of this extended component. Nevertheless, we can still determine lower and upper limits to its size considering that it is fully resolved at 8 $\mu$m with the 13 m baseline ($\sim$150~mas for a uniform disk model) and smaller than the VLTI/MIDI interferometric field of view ($\sim$250~mas at 10$\mu$m).

On the other side, the relative flux of the two components and the minor and major axis of the compact component are well constrained and do not depend on the geometry used to model the extended component in the 8-12.5 $\mu$m range. We have shown that we cannot constrain the geometry of the extended component; consequently, in the following, we will discuss only results concerning the best fitted G1 model. The wavelength dependence of F$_1$, a$_1$, a$_1$/b$_1$, and the reduced $\chi^2$ for the best-fitted G1 model are plotted in Fig.~\ref{models}.

This figure clearly exhibits the effect of the silicate band centered around 9.6 $\mu$m on the object geometry. The compact component extension and flattening strongly varies between 8 and 10 $\mu$m. At 8 $\mu$m this structure does not show any flattening and a$_1$ and b$_1$ are about 15-16~mas. Their respective size increases to about 22 and 28~mas at 10 $\mu$m and then remains almost constant up to 13 $\mu$m. On the other hand, the relative flux of the compact component F$_1$/F$_{tot}$ is almost constant between 8 and 10 $\mu$m and of the order of 0.75 $\pm$0.02. Beyond 10 $\mu$m it starts rising up to 0.82 $\pm$0.03 around 13 $\mu$m. 

As we already mentioned in the previous sections, the envelope extension variation in the N band can easily be explained by a silicate band and continuum effect. However, the apparent variation of its flattening remains hard to explain in the scheme of a classical dusty disk. If the disk is Keplerian and stratified, we would expect it to be flared. Remembering that larger wavelengths are probing cooler regions, and thus further from the central star, we would expect that the apparent flattening of a flared disk decreases with wavelength. Another possibility is that the disk might be created by an outflowing radiative wind. In this case, if the terminal velocity of the wind is high enough to avoid any stratification of the disk, we would expect that the flattening does not depend on the wavelength. Additional effects have to be considered to allow the disk to appear more flattened at larger distances from the central star. One solution might be the presence of a puffed-up inner rim, as already proposed for Herbig stars by Natta et al. (2001). However more measurements are needed to solve these issues.

If we consider that the 13 $\mu$m envelope is a geometrically thin equatorial disk, the flattening would only be due to its projection onto the sky plane, and we can estimate HD~62623's inclination angle to be 38$\pm$8$^o$ (using the 13$\mu$m value for the flattening). Unfortunately, the disk may not be geometrically thin and may also be flared, as we have already discussed. Consequently, we can only estimate a lower limit on this parameter of 30$^o$.

\subsection{Comparison with KECK measurements}

Using Keck I aperture masking interferometry, Bittar et al. (2001) partially resolved HD~62623's circumstellar environment in the near-infrared. They estimated the circumstellar environment extension using Gaussian disk models and found that its size was 6.8 $\pm$1~mas at 1.65 $\mu$m, 24.8 $\pm$1~mas at 2.26, and 27.3 $\pm$3~mas at 3.08 $\mu$m. Even if their estimation at 3.08 is almost two times larger than our compact component at 8 $\mu$m, one must keep in mind that they only obtained measurements for spatial frequencies less than 0.02 cycles/mas, and that we should compare it to a ``single" Gaussian estimation for our small and intermediate baselines (i.e., B$_3$, B$_5$, B$_6$, and B$_9$ from Table 3). Using these measurements, we obtain an average size of 29.9~mas at 8 $\mu$m. This value is clearly compatible with the one determined by Bittar et al. (2001) at 2.26 and 3.08 $\mu$m. Moreover, the estimated size does not strongly vary between 2.26 and 8 $\mu$m (i.e., only 1.21 times larger), which might indicate that the measured extension in the K band is close to the diameter of the inner rim of the dusty disk.

The existence of the two components in our HD~62623 modeling is probably due to the presence of a bright inner rim in the dusty disk which probably lies within the compact component of our model and is ``blurred" as we do not fully resolved it. Thus, the inner radius of the dusty disc should be smaller than the Gaussian equivalent size measurement of the compact component at 8 $\mu$m. 

We obtained 15~mas~$\sim$~16~R$_\star$ at 8 $\mu$m for the compact component Gaussian equivalent FWHM, and the extension at 2.26 $\mu$m is 1.21~times smaller than the one at 8 $\mu$m. Moreover, when the envelope is not fully resolved, estimating its size with a ring diameter corresponds to a value 1.24 times smaller than if we were using a Gaussian FWHM. Thus, we can roughly estimate the dusty disk inner rim radius to be of the order of R$_{in}\sim$10R$_\star$  or 3 AU.

\section{MC3D simulations of HD~62623}

\begin{figure}[t]
\centering   
\includegraphics[width=0.48\textwidth]{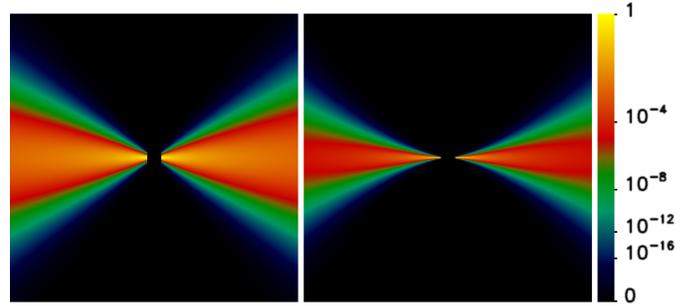}
\caption{Typical radial and vertical density distribution for an equatorial wind (left) and a Keplerian viscous disk (right). }
\label{density}
\end{figure}

\subsection{The Monte-Carlo code}

MC3D (Wolf et al. 2003) is a three-dimensional continuum radiative transfer code. It is based on the Monte-Carlo method and solves the radiative transfer equations self-consistently. Starting with a spatial density distribution of scatterers and absorbers and some primary radiative source(s), it can be used to determine the temperature in this medium by an iterative process of heating, absorption, and reemission by dust grains. When the self-consistent temperature distribution is reached, the software can compute the resulting observables: SED, wavelength-dependent images, and polarization maps.

The software allows various geometries for the density distribution : one-dimensional (i.e., dust shells) , two-dimensional (i.e, axi-symmetric disk or torus), or fully three-dimensional in spherical, cylindrical, or Cartesian coordinates (for example, to model inhomogeneities in a circumstellar disk).

\begin{figure*}[!t]
\centering   
\includegraphics[width=0.85\textwidth]{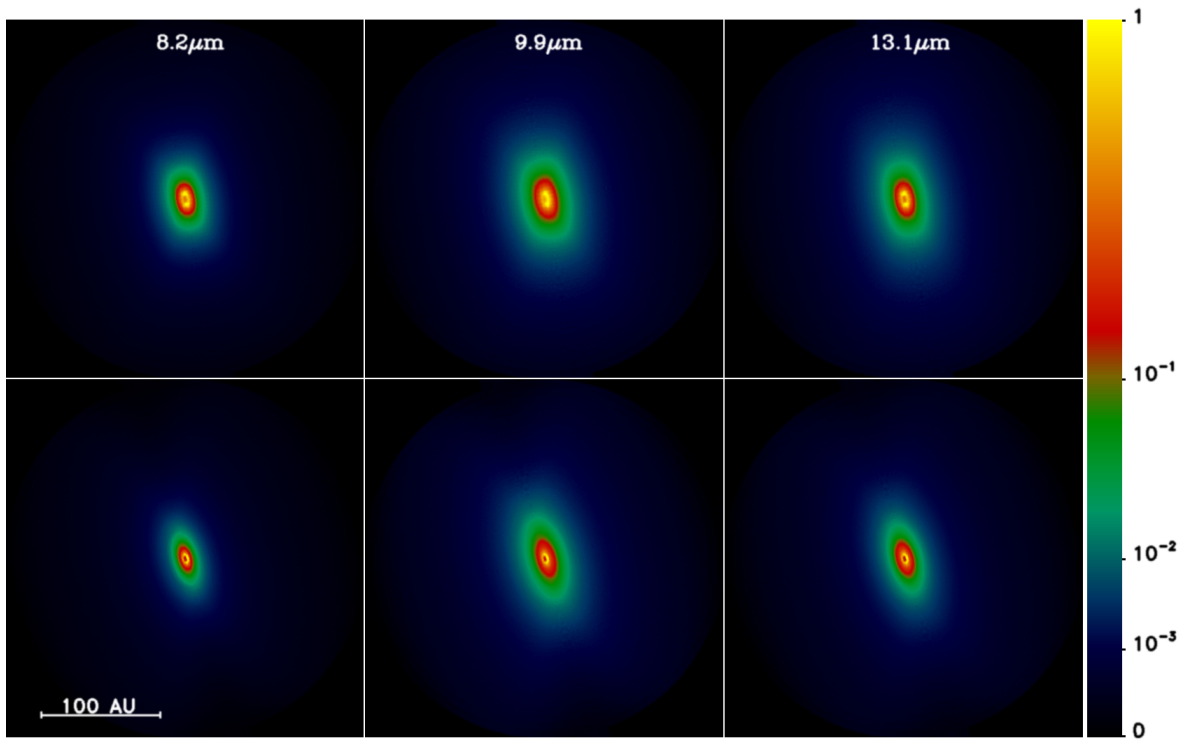}
\caption{From left to right : 8.2, 9.9, and 13.1 $\mu$m MC3D intensity maps for the best-fitted equatorial wind model (top) and Keplerian disk (bottom) of HD62623's circumstellar environment.}
\label{mc3d_maps}
\end{figure*}

\subsection{Dust disk model}

We decided to model HD~62623 with a two-dimensional dust density distribution and a radial and vertical structure, as described in Shakura \& Sunyaev (1973) and Wood et al. (2002)~:

\begin{equation}
\rm \rho(r,z) = \rho_{0}\Biggl(\frac{R_{\star}}{r}\Biggr)^{\alpha} exp\Biggl(-\frac{z^2}{2h(r)^2}\Biggl) 
\end{equation}

where $r$ is the radial distance in the midplane of the disk, $\alpha$ the density parameter in the midplane, and $h(r)$ is given by~:
   
\begin{equation}  
\rm h(r)=h_{0}\Biggl(\frac{r}{R_{\star}}\Biggr)^{\beta}
\end{equation}
         
where $\beta$ is the vertical density parameter, and h$_{0}$ is the theoretical disk scale height at R$_{\star}$. In order to be more understandable, the scale height at 100 AU (h$_{100}$) is used as an MC3D input parameter instead of h$_{0}$. The three geometrical parameters $\alpha$, $\beta$, and h$_0$ are related to the physical nature of the dusty disk. Porter (2003) shows that both the viscous disk and the equatorial wind model can theoretically allow the formation of dust around supergiant B[e] stars. A Keplerian viscous disk will imply $\alpha$=3.5, $\beta$=1.5, and h$_0$=c$_s$/$\sqrt{GM_\star}$, where c$_s$ is the sound speed. On the other hand, $\alpha$=2 and $\beta$=1 for a wind model with a constant expansion velocity. In this case, h$_0$ roughly defines the opening angle of the dusty disk. A typical density distribution for an equatorial wind and a Keplerian viscous disk are plotted in Fig~\ref{density}.

Details of the spatial distribution of chemical compositions, shapes, sizes, and other parameters of real dust grains in this object are poorly known. Large uncertainties of the real properties of dust enormously increase the free parameter space of the models to be explored. Since our purpose is to reconstruct global properties of the disk, we decided to fix the dust grain properties. We have only treated the transfer of dust radiation in this work. The gas component, present in the model only implicitly, is described by a dust-to-gas ratio of 1$\%$ within the dense disk. 

We assumed the standard interstellar size distribution from Mathis et al. (1977):

\begin{equation}
\rm \frac{\partial n(a)}{\partial a}\  \alpha\  a^{-3.5}$$ 
\end{equation}
 
where $a$ is the dust grain radius extending from $0.005$ to $0.25 \mu$m (Draine \& Lee 1984). We considered the dust grains to be homogeneous spheres.

\subsection{Model fitting}

We calculated several hundreds of models to explore the free-parameters space and we tried to constrain their values and uncertainties. We focused on both the viscous disk and the equatorial wind models defined in the previous section to test if we can clearly discriminate between them. Finally, we also tried many intermediate models which imply that the disk is slowly expanding and not perfectly stratified.

We considered the following outputs from our simulations:
\begin{enumerate}
\item The total SED computed from 1 to 20 $\mu$m that can be directly compared to our reconstructed SED, including our MIDI data presented in Sect. 2.
\item Twenty 301x301 pixels intensity maps in the MIDI wavelength range, i.e, 8-13.5~$\mu$m, with a pixel size of 0.7 mas. Each image allows us to fit the spectrally dispersed visibilities at one wavelength. The transformation of the images into a visibility signal is straightforward: the images are rotated by the PA angle of the object onto the sky and then collapsed as 1D flux distribution in the direction of each of the nine baselines. These vectors are Fourier transformed and normalized.
\end{enumerate}

For each model we compute the SED and visibilities $\chi^2$ separately. The use of a single global $\chi^2$ would have implied solving the problem of the relative weight of the SED and visibility in the computation. Finally, we decided to concentrate our efforts on the visibility fit since it contains more information on HD~62623's circumstellar environment geometry than the SED. Nevertheless, for each model we verify that the fit of the SED is accurate enough, both in term of IR-excess and silicate band emission intensity.

\begin{table}[!b]
\centering
\begin{tabular}{ccc}
Parameter & Best Wind Model & Best Keplerian disk model\\
\hline
\hline
R$_{in}$&4$\pm$0.5AU&3.7$\pm$0.5AU\\
R$_{out}$&150AU&150AU\\
M$_{\rm env}$& 2.10$^{-7}$M$_\odot$& 2.10$^{-7}$M$_\odot$\\
T$_{\rm dust}$&1250K&1250K\\
$\alpha$&2&3.5\\
$\beta$&1&1.5\\
h$_{\rm 100 AU}$&24AU&28AU\\
Inclination angle&60$\pm$10$^o$&60$\pm$10$^o$\\
Major-axis P.A.&10$\pm$10$^o$&15$\pm$10$^o$\\
\end{tabular}
\caption{Parameters of the best fitted MC3D equatorial wind and Keplerian disk model. For both models T$_{\rm eff}$=8250K and L$_\star$=17000L$_\odot$ (i.e., corresponding to R$_\star$=65R$_\odot$) and d=650pc.\label{MC3D_fit}}
\end{table}

\begin{figure*}[t]
\centering   
\includegraphics[width=0.46\textwidth]{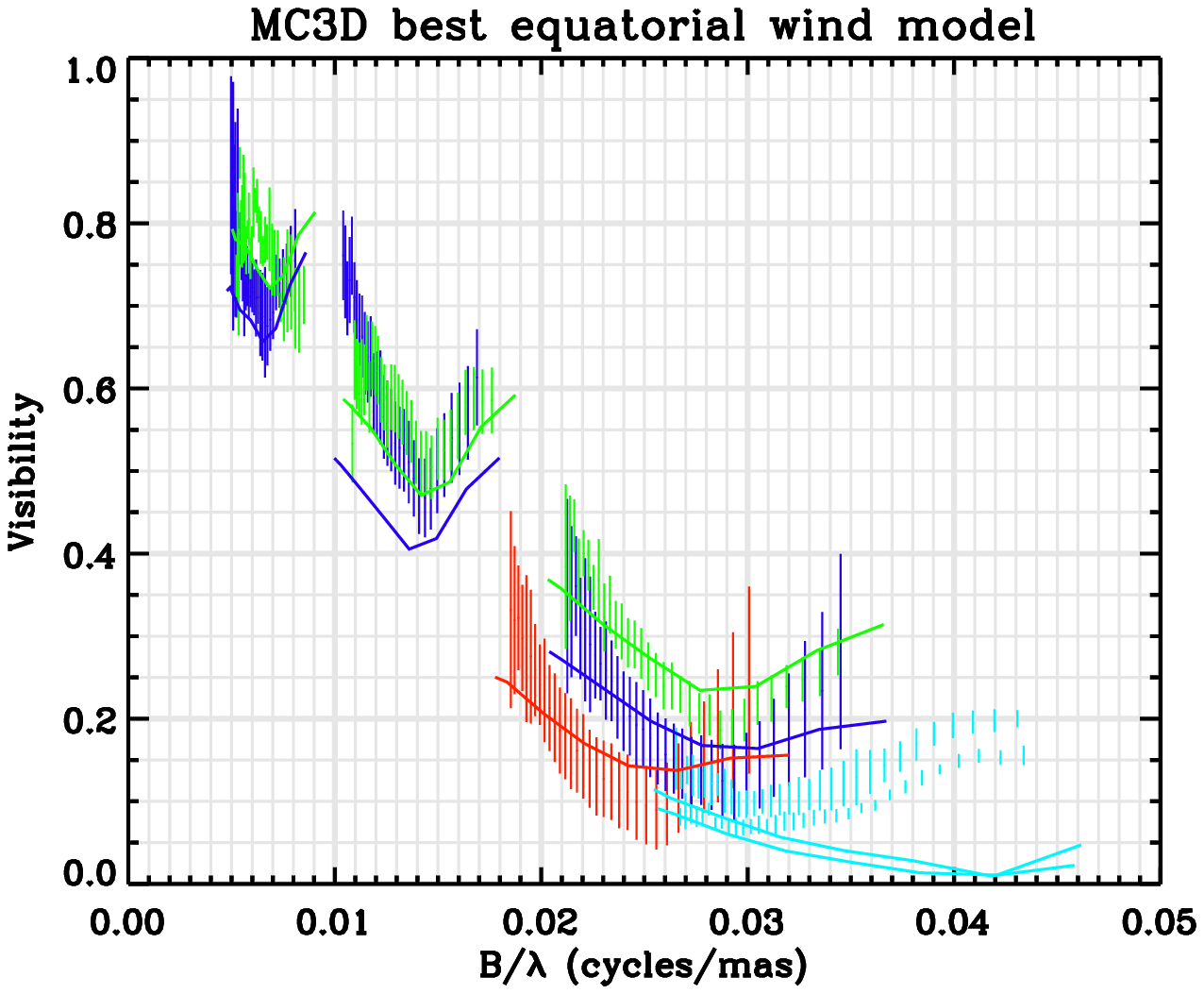}
\includegraphics[width=0.46\textwidth]{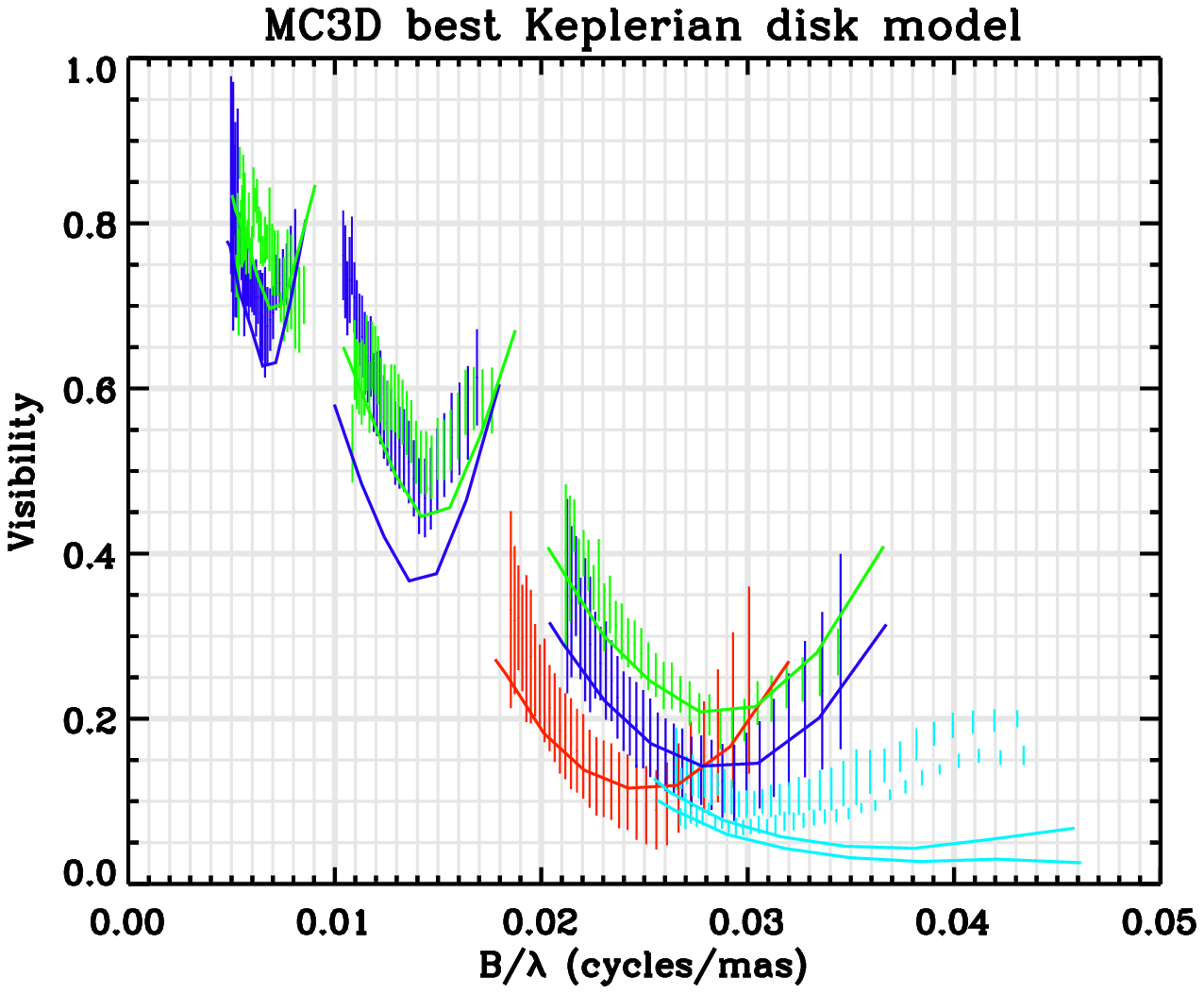}
\includegraphics[width=0.46\textwidth]{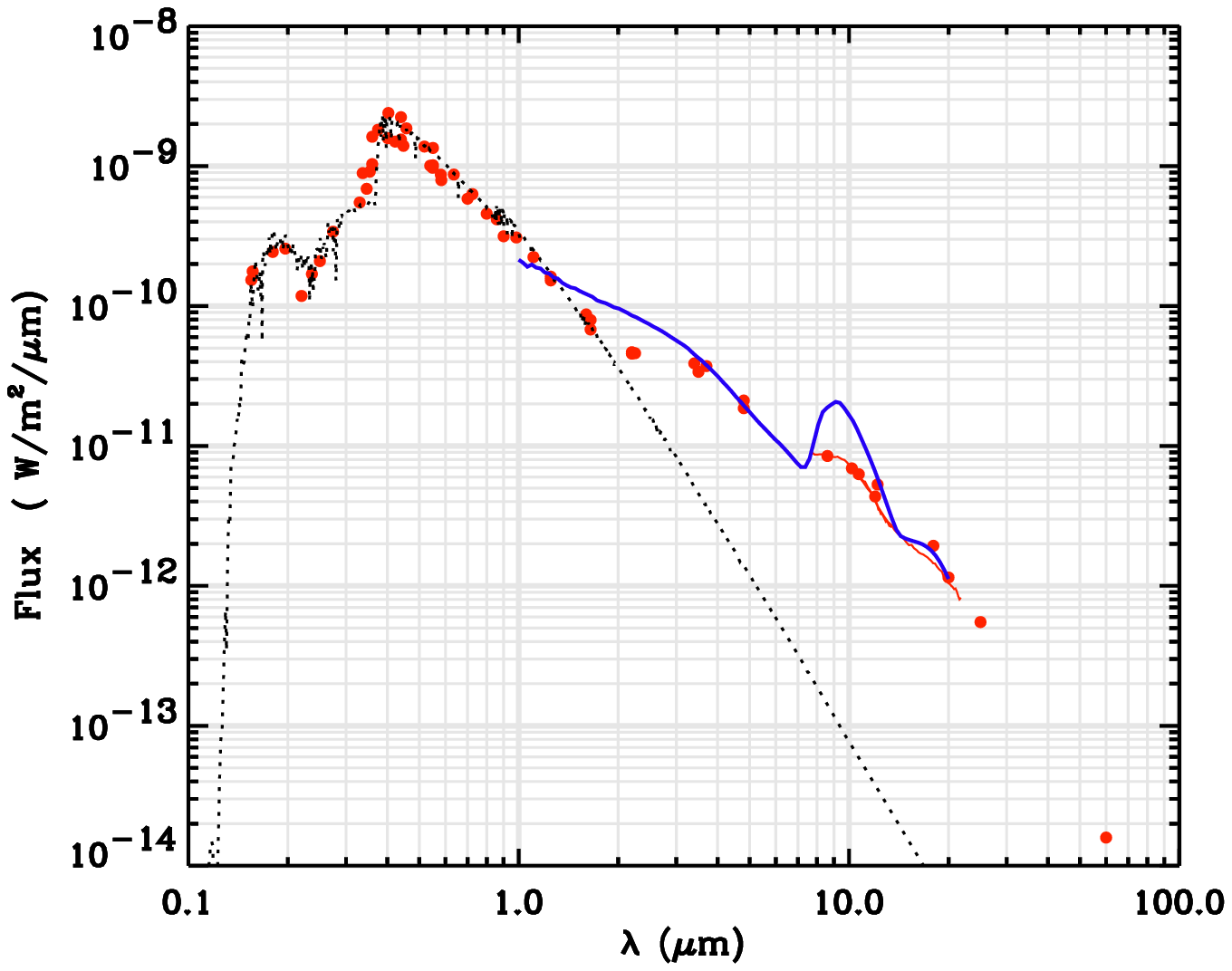}
\includegraphics[width=0.46\textwidth]{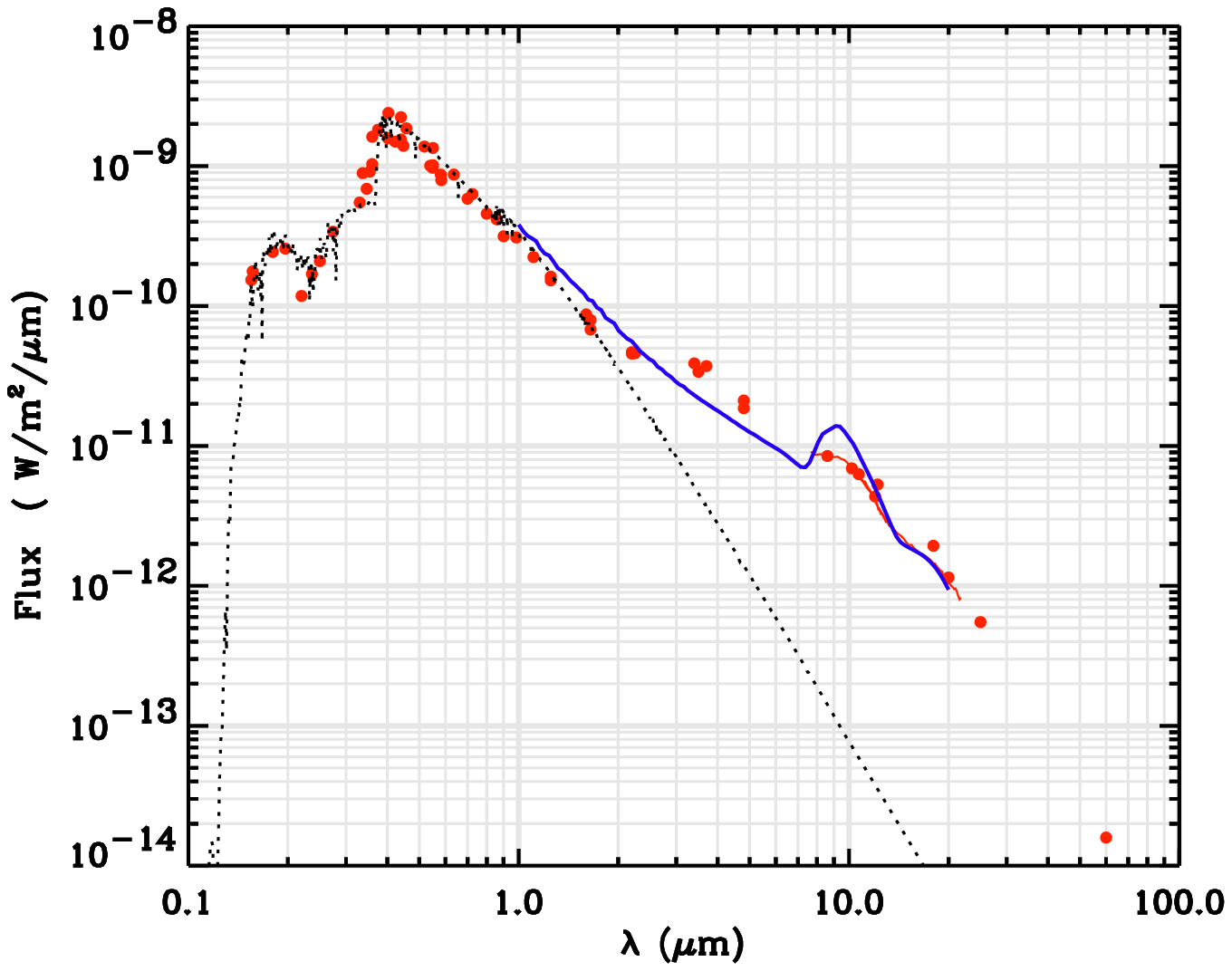}
\caption{Fit of the nine visibility measurements (top) and SED (bottom) of the best equatorial wind model (left) and Keplerian viscous disk model (right). The dotted lines correspond to the models and the errors bars to the data.  }
\label{mc3d_fit}
\end{figure*}

\subsection{The best models}

The parameters of the best equatorial wind and Keplerian viscous disk models are presented in Table~\ref{MC3D_fit}, and the corresponding 8.2, 9.9, and 13.1$\mu$m images for both geometries are plotted in Fig~\ref{mc3d_maps}. The fit of the visibilities  is plotted in Fig~\ref{mc3d_fit}. These figures show that both scenarios are able to reproduce the general trends of the seven shorter baseline visibility variations, but the ones measured for B$_7$ and B$_8$ are significantly underestimated for wavelengths shorter than 11 $\mu$m. Such a residual visibility might suggest the presence of structures not fully resolved with the 70~m baselines, which are not taken into account in our MC3D model. Some hypotheses on the nature of these structures will be discussed in Sect. 6.1. 

The reduced $\chi^2$ on the interferometric data of the best equatorial wind and Keplerian disk are 14.9 and 11.5, respectively. These high values are mainly due to the poor fit of the 70m baseline. Thus, skipping B$_7$ and B$_8$ measurements, the reduced $\chi^2$ reach values of 2.8 and 2.4, respectively. The main qualitative difference between the two scenarios is the silicate band visibility drop scale. This favors the hypothesis of the stratification of the disk.

\begin{figure*}[!t]
\centering   
\includegraphics[height=0.275\textheight]{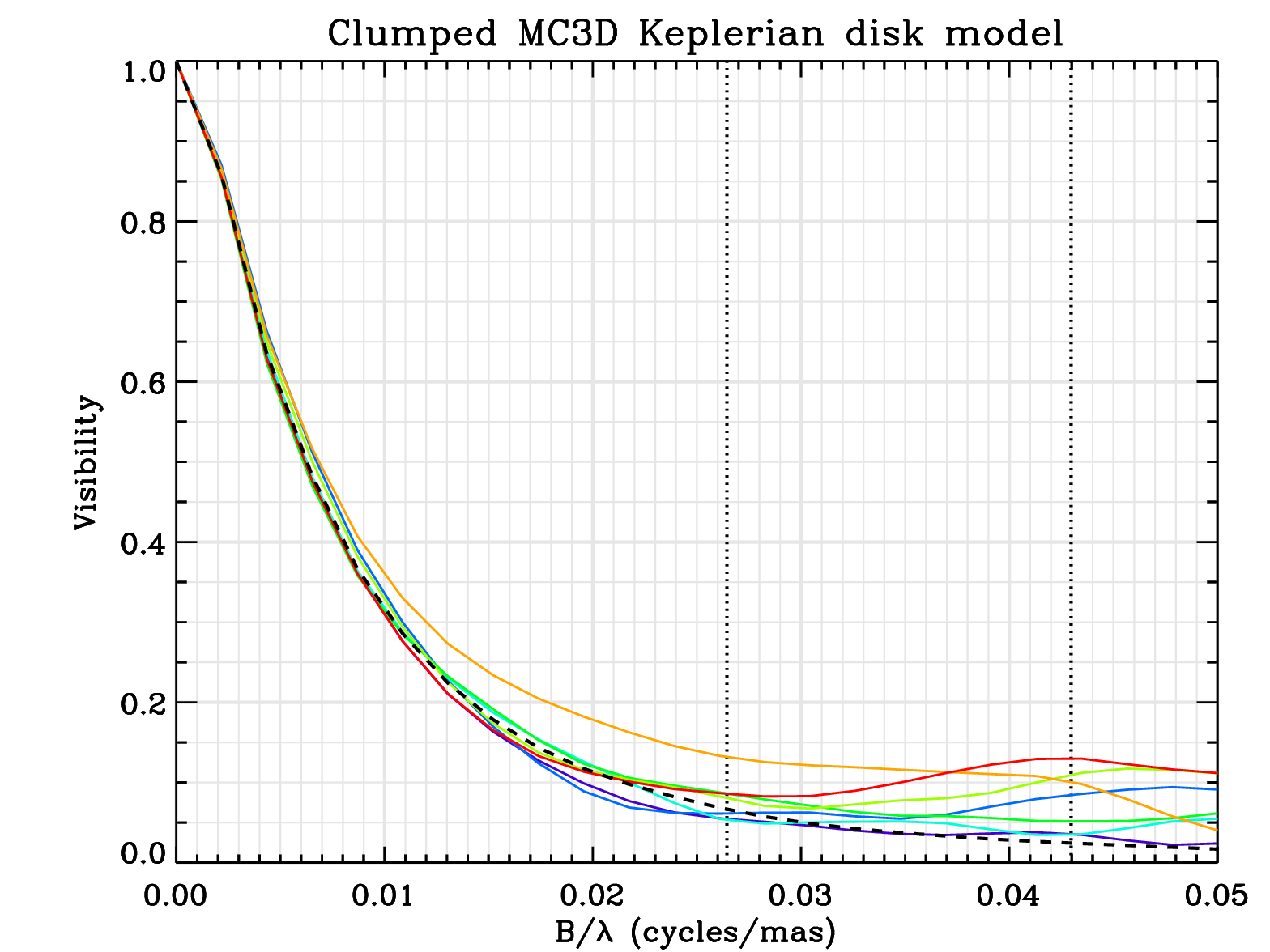}
\includegraphics[height=0.26\textheight]{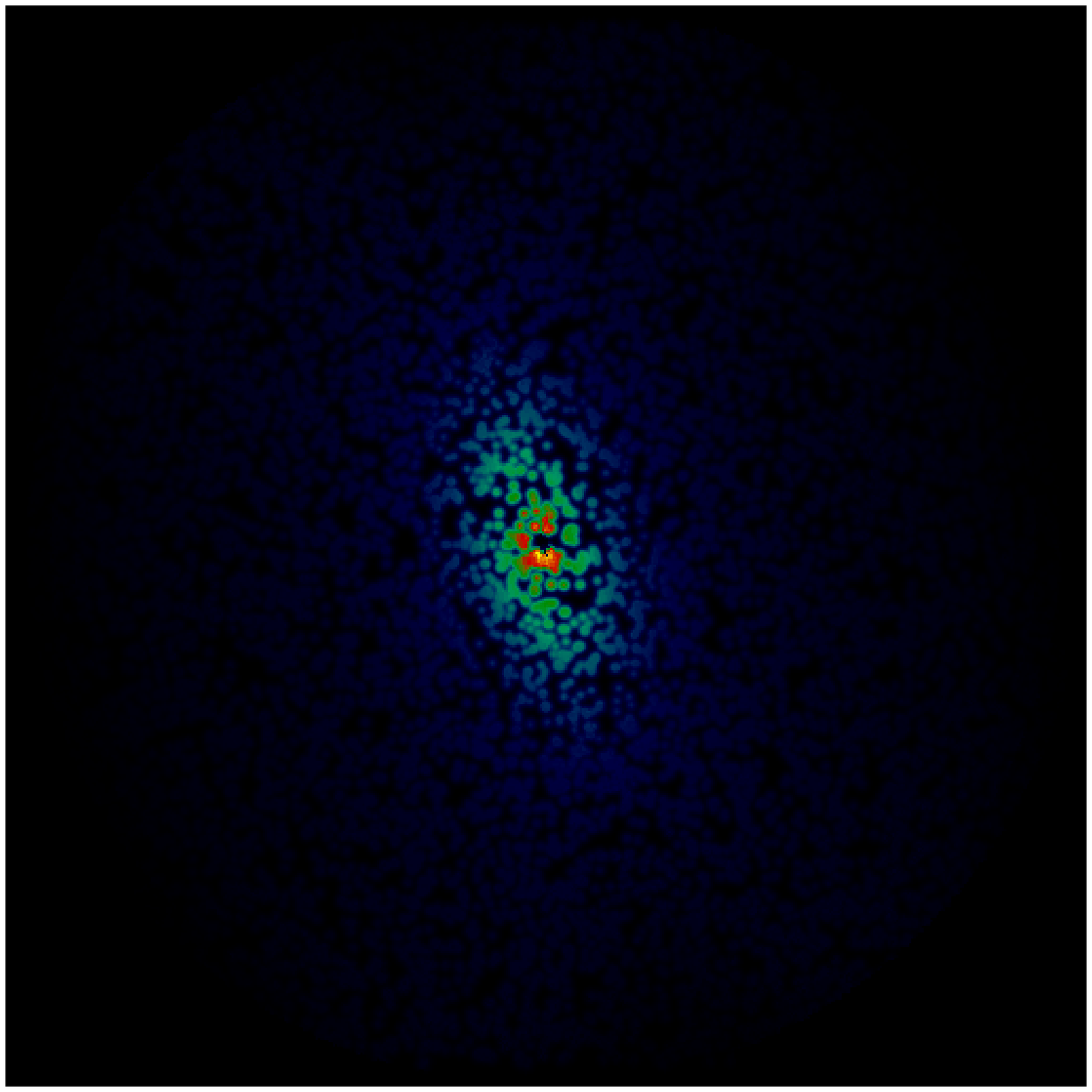}
\caption{Left : Visibility plotted as a function of the spatial frequency for several random clump distributions  (colored solid lines) with N$_{clump}$=10000 and FWHM$_{clump}$=3~mas, using the 8 $\mu$m intensity map from our best MC3D wind model as an input. The visibility corresponding to this map is also plotted as a dashed line for comparison. The vertical dotted lines represent the position of the spatial frequencies for a 70~m baseline at 8 and 13 $\mu$m. Right : Example of a clumped map with a random distribution for the same parameters.}
\label{clump}
\end{figure*}

However, the differences are small, and a large set of intermediate values of $\alpha$ and $\beta$ are also possible so that the disk stratification cannot be inferred accurately. Actually, as can be seen in Fig~\ref{mc3d_maps}, the geometrical differences between these scenarios are too small to be properly constrained by our nine VLTI/MIDI measurements with uncertainties of the order of 5$\%$. 

Nevertheless, despite our inability to deduce accurate values of the radial and vertical disk structure from the visibility fits, we clearly constrain the inner disk radius, R$_{in}$=3.85$\pm$0.7 (or R$_{in}$=4$\pm$0.5 AU for the equatorial wind and R$_{in}$=3.7 $\pm$0.5 AU for the stratified disk). This represents the first measurement of the extension of the inner rim of a B[e] supergiant dusty disk. The mass of the envelope does not strongly depend on the disk stratification and is also well constrained, i.e., 2 $\pm$110$^{-7}$ M$_\odot$, so is the inclination angle, i.e., 60 $\pm$10$^o$. Finally, the position of the major-axis that is roughly perpendicular to Yudin \& Evans (1998) polarization angle measurement 95$^o$.

The SED fit for each scenario is presented in Fig~\ref{mc3d_fit}. For both scenarios, the fit of the SED is not good enough, especially the 9.6$\mu$m silicate peak, which is too strong for both models, whereas the IR-excess is overestimated for the equatorial wind and underestimated for the Keplerian disk scenario. However, the general agreement is better for the Keplerian disk, and we note that both models managed to reproduce the 13-20 $\mu$m SED well.

In fact, an accurate fit of the silicate feature is very often difficult to obtain. The strength of the feature depends on the size and composition of the silicate grains, but also on the level of the underlying continuum, thus, on the quality of the model density structure. The near-IR region of the SED is also strongly influenced by the detailed structure of the inner rim of the disk which, in this study, is assumed to be a wall. Recently, many studies were devoted to this point in the field of young stellar objects, involving a puffed-up inner rim and dust settling (Dullemond et al. 2001, 2004; Tannirkulam et al. 2007).

For the Keplerian disk, the near-IR flux is underestimated because the inner rim is probably not large enough. On the contrary, the equatorial wind model overestimates the near-IR as a direct consequence of the smaller density coefficient $\alpha$, making the inner rim almost as opaque as a true wall. Moreover, the flaring of the density structure is also larger; thus, more of the surface of the dust disk is directly exposed to the stellar radiation and the vertical extent of the structure is probably over-estimated.

\section{Discussions}

\subsection{Nature of the residual visibility}

As already mentioned in Sect. 4.1, the fact that the two largest baselines, i.e. B$_7$ and B$_8$, do not exhibit the same wavelength dependence as the shorter ones, in particular the lack of a clear drop of the visibility in the 9.6 $\mu$m silicate band, might be a clue that both the continuum and the silicate emissions are over-resolved with these baselines. Since the visibility does not reach zero with these baselines but 0.17$\pm$0.05 at 8 $\mu$m and 0.12 $\pm$0.03 at 13 $\mu$m, this might advocate for the presence of a unresolved structure which contributes to the total flux at least to 17$\%$ at 8 $\mu$m and 12$\%$ at 13 $\mu$m, assuming it is fully unresolved.

This assumption is confirmed by our MC3D models that clearly show the impossibility to reproduce the B$_7$ and B$_8$ measurements simultaneously to the other ones. Actually, with these 70~m baselines, the dusty disk inner rim (i.e., R$_{\rm in}$ = 3.85 AU = 12.8R$_\star$) is almost fully resolved. Thus, the visibility should be close to zero. Taking a smaller R$_{\rm in}$ would lead to a better fit of B$_7$ and B$_8$ visibilities, but to an overestimation for the other ones.  Moreover, the fit of the largest baselines would not be totally satisfying since the silicate band effect would be clearly visible in the modeling.

Thus, in the following, we investigate two different hypotheses to explain the residual visibility measured with the 70~m baselines.

\begin{figure*}[!t]
\centering   
\includegraphics[height=0.28\textheight]{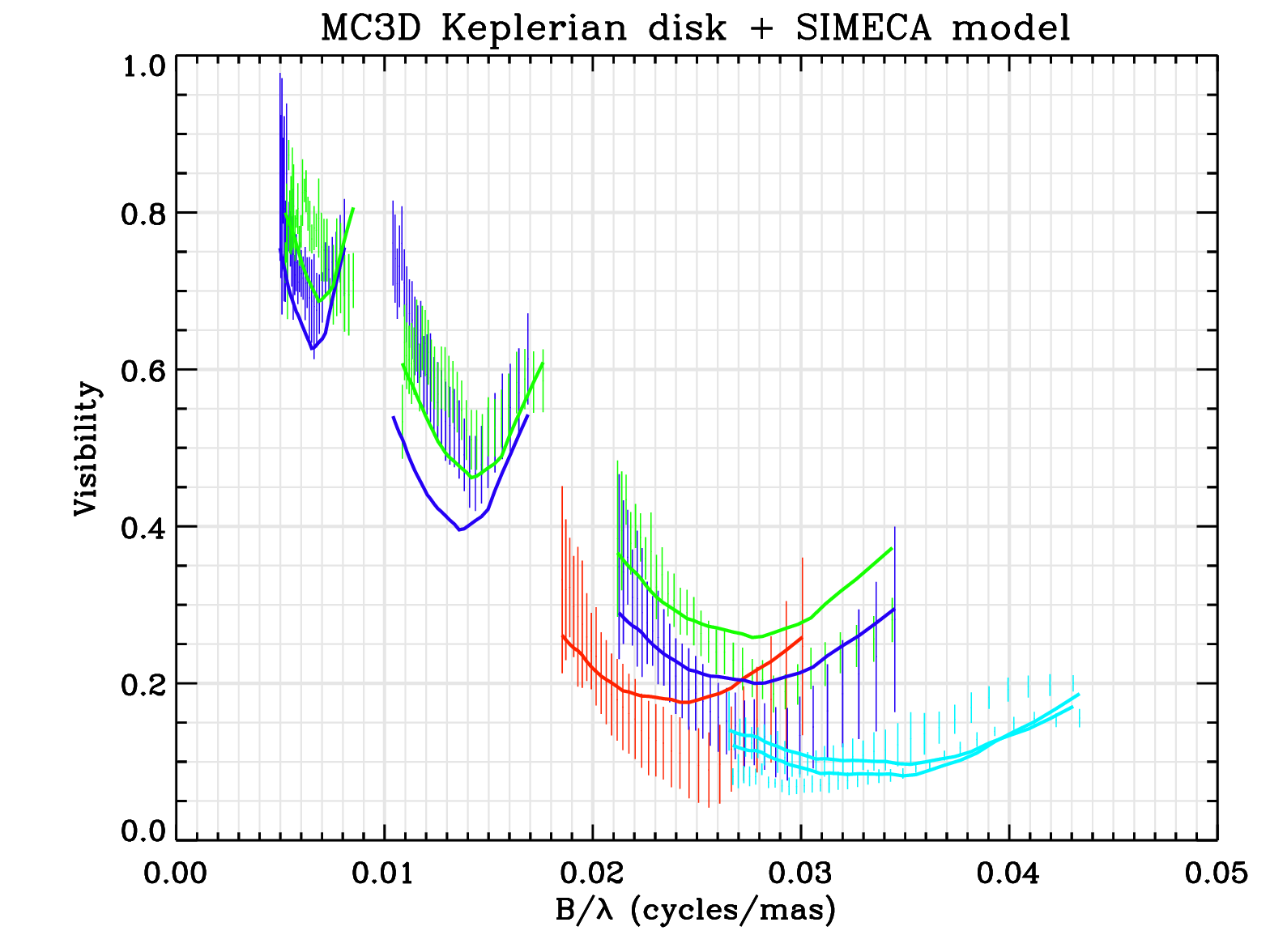}
\includegraphics[height=0.27\textheight]{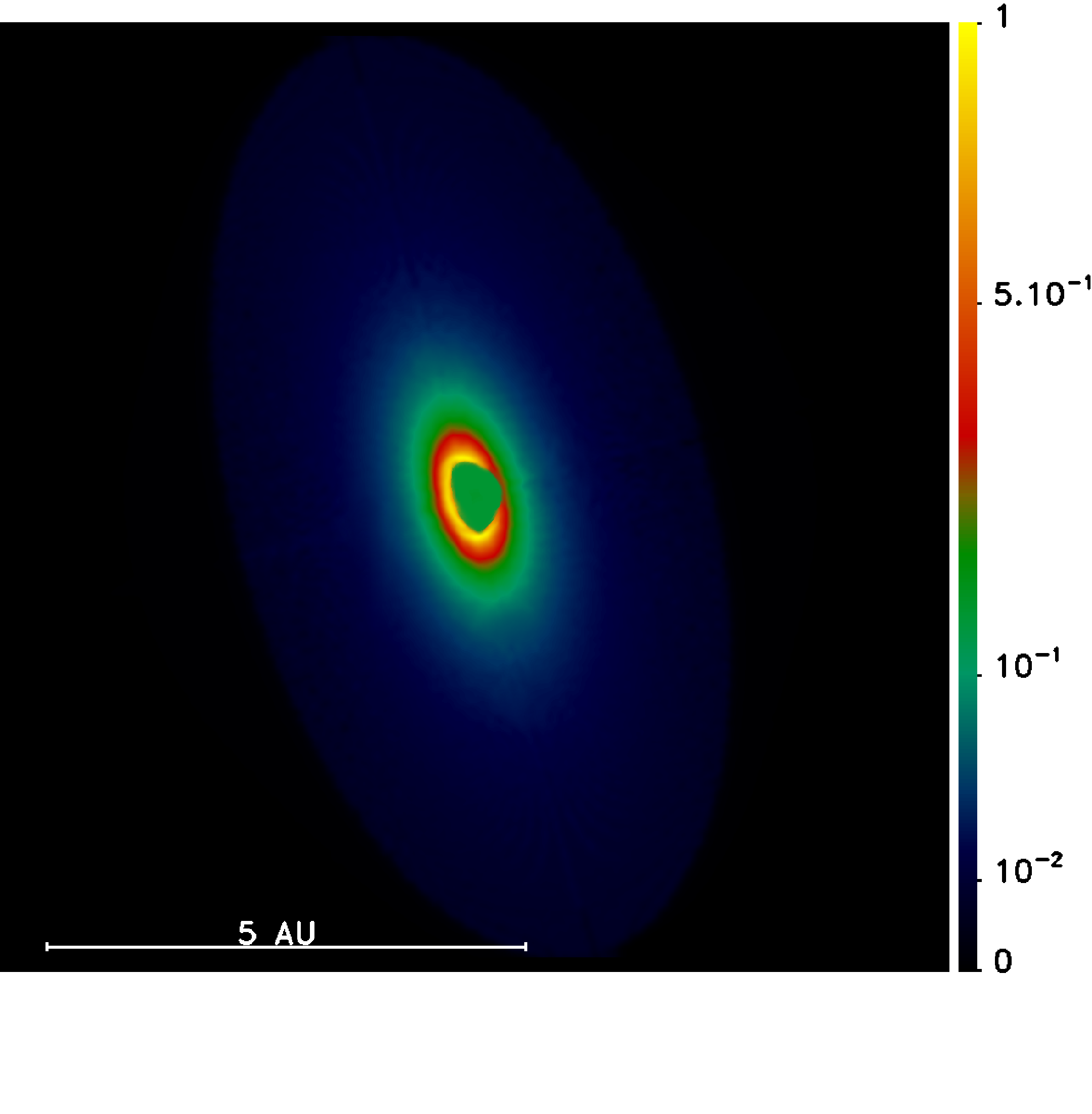}
\caption{Left : Visibility plotted as a function of the spatial frequency for the SIMECA + MC3D model with parameters from Table~\ref{SIMECA_table}. Right : corresponding 8$\mu$m SIMECA intensity map.}
\label{SIMECA}
\end{figure*}

\subsubsection{Clumping of the dust disk}

Effects of clumping on the visibility are not trivial and depend mainly on the number of clumps and their extension. If they are unresolved, i.e. D$_{clump}\ll$~B/$\lambda$, they will significantly increase the visibility level. However, if their number is small, they will act like multiple objects and produce oscillations structures in the visibility function as a function of the spatial frequency. Consequently, since a clear increase of the visibility level but no oscillations are evident in our data, we can assume that the clumps are unresolved and numerous. 

To simulate the effect of clumping on our visibility measurements, we use a simple and ad-hoc method to produce a clumped, dusty disk emission map. We compute a random clumps map in which each clump is a normalized Gaussian. We then multiply this map by an MC3D emission map. The resulting map consists of a clumped version of our previous MC3D emission map. Fig~\ref{clump} (left) presents the visibility plotted as a function of the spatial frequency for several random clump distributions using a 13.1 $\mu$m intensity map from our best MC3D Keplerian disk model as an input and with N$_{clump}$~=~10000 and FWHM$_{clump}$~=~3~mas. An illustration of a randomly clumped map is also plotted in Fig~\ref{clump}~(right). 

This figure exhibits the random clump distribution effect on the visibility. For the selected clump parameters, it clearly shows that when the global disk structure is fully resolved (i.e. B/$\lambda>$0.05), the visibility can take any value between 0 and 0.2, depending on the clump distribution. Such random effect due to clumpiness have already been studied using fully radiative transfer clumped-disk models by H\"onig et al. (2006) for AGN tori. 

Finally, it seems that random clump distributions with the previous parameters can produce visibilities compatible with our measurements. We note that this result is strongly dependent on the baseline P.A., and two baselines separated by a few degrees can produce totally different visibility profiles. Consequently, knowing that our two largest baselines are separated by nearly 20$^o$, the fact that they exhibit nearly the same trend can be considered an argument against the dust disk clumping hypothesis. However, we cannot totally exclude that a random distribution of clumps may reproduce the observed visibilities.

\begin{table}[!b]
\centering
\begin{tabular}{ccc}
\multicolumn{2}{c}{Parameter} & Value\\
\hline
\hline
$\rho_0$     			 & Photospheric density         	& 4x10$^{-12}$ g.cm$^{-3}$ \\
$\phi$(pole) 		 	& Polar mass flux             	  &10$^{-11}$ M$_\odot$.yr$^{-1}$\\
$\phi$(eq)   		 	& Equatorial mass flux         		&10$^{-10}$ M$_\odot$.yr$^{-1}$\\
m$_1$							&	Mass flux law exponent 					& 100\\
v$_\infty(pole)$	& Polar terminal velocity 				& 200 km.s$^{-1}$\\\
v$_\infty$(eq)		& Equatorial terminal velocity 		&0.1 km.s$^{-1}$\\
m$_2$							&	Terminal velocity law exponent  &100\\
i            			& Inclination angle           	  &60$^o$\\
P.A.         			& Major-axis orientation      	  &15$^o$\\
T$_{eff}$					& Effective Temperature						&8250 K\\
R$_\star$					&	Stellar radius									& 65 R$_\odot$\\
d									& Distance												& 650 pc\\
\end{tabular}
\caption{Parameter values for the SIMECA model.\label{SIMECA_table}}
\end{table}

\subsubsection{Inner gaseous disk}

To test if the inner gaseous disk of HD~62623 can be responsible for the residual visibility, we used the SIMECA code developed by Stee (1996) to model the circumstellar gaseous environment of active hot stars. This code is based on a latitudinal-dependent CAK wind model (Castor, Abbott, \& Klein 1975) and large velocity gradient radiative transfer (Sobolev 1960) and is able to compute SED, hydrogen emission line profiles, and intensity maps in the continuum and in the emission lines. However, in our case we only need to compute radiative transfer in the continuum in order to obtain 8 and 13 $\mu$m intensity maps and 0.1-100 $\mu$m SED. The aim of this modeling is not to constrain the physical properties of HD~62623's gaseous inner envelope, which is not possible due to our small number of measurements, but rather to determine if this component can be responsible for the 70~m N-band residual visibility. 

\textbf{In the SIMECA code,  m$_1$ and m$_2$ are exponent used to set the latitudinal dependence of the mass-flux ($\Phi$) and terminal velocity (v$_\infty$), respectively:}

\begin{equation}
\Phi(\theta) = \Phi(pole)+\bigl(\Phi(eq)-\Phi(pole)\bigr) sin^{m_1}\theta
\end{equation}
\begin{equation}
v_\infty(\theta) = v_\infty(pole)+(v_\infty(eq)-v_\infty(pole)) sin^{m_2}\theta
\end{equation}

Typical values of the terminal velocity of an A supergiant wind are of the order of 500 kms$^{-1}$ (Achmad et al. 1997) although Lamer et al. (1995) derived a terminal velocity of 150 $\pm$50 kms$^{-1}$ for HD~62623 from IUE observations. However, HD~62623 may be rotating at a significant fraction of its breakup velocity; i.e., V~sin~i~$\sim$~70kms$^{-1}$ and 160~kms$^{-1}<~$v$_c~<$~250 kms$^{-1}$, depending on the value of M$_\star$. Thus, the expansion velocity at the equator can be smaller, as predicted by the bi-stability model (Lamers \& Pauldrach 1991). Studying the classical Be stars $\alpha$ Arae and $\kappa$ CMa with the VLTI/AMBER instrument, Meilland et al. (2007a, 2007b) found that equatorial expansion velocities were smaller than 10 kms$^{-1}$. Moreover, a spectroscopic follow-up of the variable Be stars $\delta$ Sco (Miroshnichenko et al. 2003) and Achernar (Kanaan et al. 2007) leads to values of the order of  0.2-0.4 kms$^{-1}$. 

Thus, we have tested various geometries and kinematics to simulate HD62623's gaseous environment~: a spherical wind with terminal velocities between 10 and 500kms$^{-1}$, and a flattened envelope with terminal velocity between 100-500 kms$^{-1}$ at the pole and 1-100 kms$^{-1}$ at the equator. We have also tested various mass-loss rates and densities. This modeling provides a typical extension for the gaseous environment of 5-10~mas in the N-band (assuming a 65R$_\odot$ star located at 650 pc). Thus, the visibility of such a component for a 70~m baseline is about 0.8-0.95 at 8 $\mu$m and 0.91-0.97 at 13 $\mu$m. Finally, this inner gaseous disk should be almost non-resolved and can explain the observed residual visibility. 

Assuming that all other components are fully resolved, we can roughly deduce the 8 and 13 $\mu$m flux of this structure to be 20$\%$ and 12$\%$ of the total flux, respectively. To produce such a contribution to the observed flux, the mass of the gaseous envelope has to be of the order of 4.10$^{-7}$ M$_\odot$. 

To illustrate this scenario we combine a MC3D and a SIMECA model without taking into account the interaction between the gas and the dust. We calculate the complex visibility for each contribution separately and finally add both components using the formula~:

\begin{equation}
V_{tot}= \frac {V_1 F_1 + V_2 F_2} {F_{tot}}
\label{Eq.1}
\end{equation}

\begin{figure*}[!t]
\centering   
\includegraphics[width=0.95\textwidth]{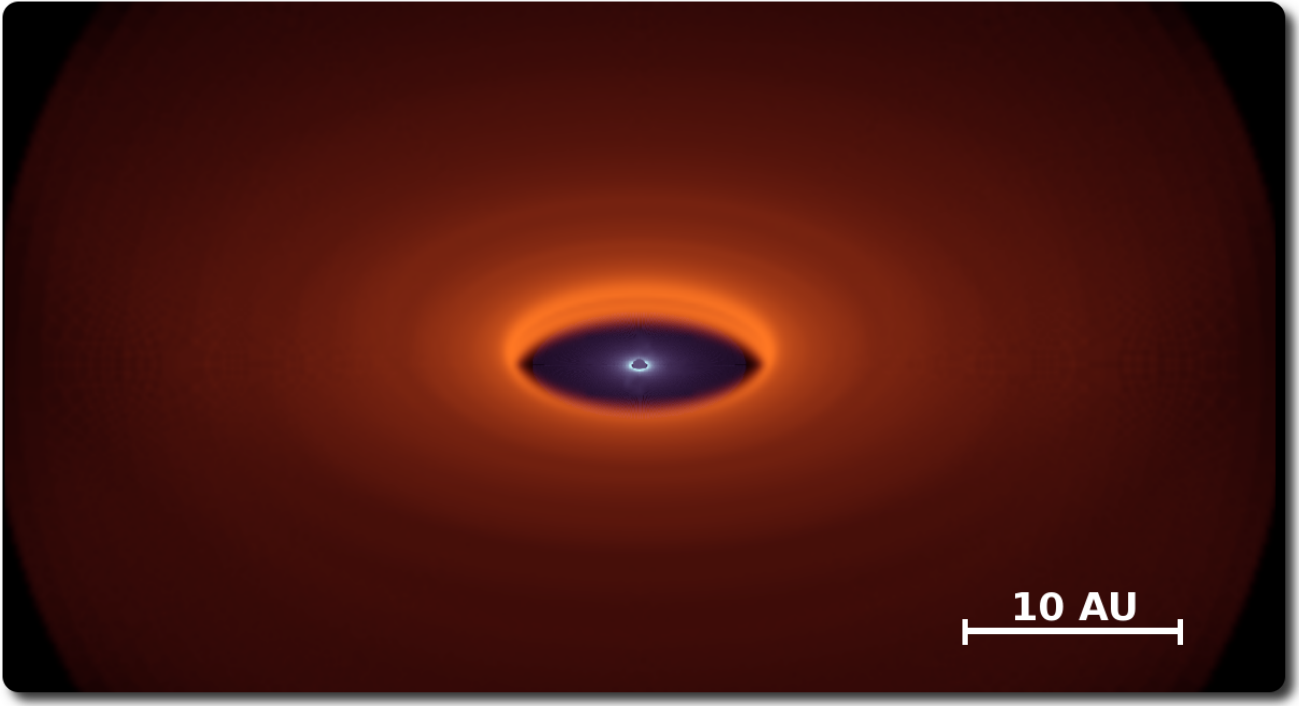}

\caption{A composite map of HD~62623's circumstellar environment in the N band from our MC3D + SIMECA model.}
\label{final}
\end{figure*}

The resulting visibilities are plotted as a function of the spatial frequency in Fig~\ref{SIMECA}. An 8 $\mu$m SIMECA intensity map for the corresponding model is also plotted in this figure. The chosen density distribution is highly flattened in order to reproduce the visibility difference between B$_7$ and B$_8$. To obtain a good fit of the interferometric measurements we need to slightly change the values of our MC3D best model since the compact gaseous emission increases the global visibility level. Thus, by increasing both the total mass to 3.5.10$^{-7}$ M$_\odot$ and the inner dust radius to 4.5 we managed to 
obtain a satisfying fit with a total reduce $\chi^2$ for the nine baselines of 4.6. Table~\ref{SIMECA_table} present values of the SIMECA parameters corresponding to this model.  Details on the physical meaning of these parameters are presented in Stee et al. (1996). Finally, Fig~\ref{final} presents a composite image of the MC3D + SIMECA model close to the central star. We note that the gaseous envelope is not strongly constrained and that these parameters have to be considered only as an example of a relatively ``good" model.

\subsection{Physical conditions in the inner gaseous envelope}

One of the main issue regarding supergiants showing the B[e] phenomenon is how dust can form in the surroundings of these highly luminous, hot stars. To form dust in a circumstellar environment, two conditions must be met: the temperature must be lower than the dust sublimation temperature (i.e., between 800 and 2000K, depending on the chemical composition of the dust), and the density of the chemical elements involved in the dust formation have to be higher than a critical value.

These two conditions are hard to find at the same location in the circumstellar environment of these stars since, for most models, the density quickly falls with the distance. However, the stellar radiation might be shielded by the gaseous inner envelope. This would allow the temperature to decrease faster than in a normal reprocessing disk, and thus, the sublimation temperature could be reached closer to the central star.

Since we managed to significantly constrain the inner rim extension of a supergiant dust disk for the first time, we can test this hypothesis by inferring the temperature variation in the inner gaseous envelope. We assume that the temperature distribution follows a power law as a function of the distance to the central star : 

\begin{equation}
\rm T(r)=T_{eff}. \Biggl(\frac{R_\star}{r}\Biggr)^\gamma
\end{equation}  

where T$_{eff}$ and R$_\star$ are the stellar effective temperature and radius. In the vacuum, the only effect on the stellar radiation is the geometrical dilution of the energy and thus $\gamma$=0.5. For a simple reprocessing disc, i.e. stellar radiations absorbed by the disk and re-emitted at the LTE temperature, $\gamma$=3/4 (Porter 2003). If the stellar radiation is highly shielded by the gaseous environment, then $\gamma\gg$3/4.  

Using R$_{in}$=3.85$\pm$0.7~AU (or 12.8$\pm$2.5$_\star$) and T$_{in}$=1250K, we can determine the value of $\gamma$ for HD~62623's circumstellar disk. Using equation (4) with r=R$_{in}$ leads to :

\begin{equation}
\gamma = \frac { ln \Bigl(\frac{T_{in}}{T_{eff}}\Bigr)}{ ln \Bigl(\frac{R_\star}{R_{in}}\Bigr)}= 0.74 \pm 0.1
\end{equation} 

This value of $\gamma$ is very close to the value of the reprocessing disk. If we use the estimation of the inner radius from our SIMECA + MC3D model, i.e. R$_{\rm in}$=4.5~AU, we find $\gamma\sim$0.79, again close to the reprocessing disk. Consequently, no strong shielding effect of the stellar radiation by the gaseous envelope is needed to obtain a temperature low enough to form dust at R$_{in}$. If the star is rotating close to its critical velocity, the equator can even be cooler thanks to the Von Zeipel effect. This would even decrease the value of the $\gamma$ parameter. 

\subsection{Comparison with other A-type supergiants}

HD~62623 is one of the coolest stars that exhibits the B[e] phenomenon. Actually, it's a member of the very rare A[e] supergiant class. A similar object  has recently been discovered in the Small Magelanic Cloud by Kraus et al. (2008). Apart from the spectral features relative to the B[e] phenomenon exhibiting the presence of a highly illuminated gas and dust region, these stars seem to share the same properties as their ``normal" counterparts.

Normal-A-Type supergiants are known to exhibit radiation-driven stellar wind with a terminal velocity of the order of several hundreds of kms$^{-1}$ and mass-loss between 10$^{-8}$M$_\odot$ and 10$^{-6}$M$_\odot$ (Achmad et al. 1997). However, no dust has been found in the circumstellar environment of these standard A-Type supergiants. As already discussed in the previous section, this implies that the critical density of chemical elements that can be involved in the dust formation (i.e., mainly Si, O, and C) is reached too close to the star, where the temperature is still higher than the sublimation limit. This is probably due to the fact that the relatively high wind terminal velocity is reached close to the central star. Aufdenberg et al. (2002) studied the mass-loss and radiative stellar wind of the A-supergiant prototype Deneb. They approximated the wind velocity assuming:

\begin{equation}
v(r) = v_\infty\bigl(1-R_\star/r\bigr)^\beta
\end{equation}

and found that v$_\infty$ = 225 kms$^{-1}$ and $\beta$ = 3.

\textbf{However, even if Deneb and HD~62623 have the same spectral class (i.e., A2Ia), their physical parameters can be different enough to affect the wind properties. For instance, Schiller \& Przybilla (2008) derived a luminosity of  1.96x10$^5$L$_\odot$ for Deneb, thus about 12 times our estimation for HD~62623. Remembering that the mass loss rate scales the luminosity, this should imply to a stronger wind for Deneb. But, in their estimation the authors used a distance of 802 pc, while the one derived from ``Hipparcos the New Reduction" (van Leeuwen, 2007) is 1.86 smaller (i.e., 432 $\pm$ 61 pc). Finally, using this latest estimation of the distance, we derived a luminosity of 5.5~$\pm$~1~x~10$^4$~L$_\odot$,  closer to  HD~62623 luminosity.}

\textbf{Thus, we can still use Deneb's wind parameters as a rough estimate for HD~62623. Consequently,} at 12.8 R$_\star$, i.e. dust sublimation radius of HD~62623 determined by our MIDI observation and MC3D modeling, the expansion velocity should be of the order of 176 kms$^{-1}$.

It is difficult to explain why HD~62623 is forming dust in its circumstellar environment while Deneb exhibits a purely gaseous stellar wind. Among the putative explanations for such a difference is a possible effect due to the stellar rotation. In the bi-stability model developed by Lamers \& Pauldrach (1991), the fast rotation induces a strong opacity change in the wind between the pole and the equator resulting in the existence of two distinct regions in the circumstellar environment; i.e., a quickly expanding, low density polar wind and a dense equatorial outflow with a smaller terminal velocity. However, they conclude that this mechanism is more efficient for stars with 15000K$~<~$T$_{eff}$$~<~$30000K. For lower temperature, the mass-loss rate is usually too low to reach the optically thick wind near the equator unless the stars are almost critical rotators. 

In the case of HD~62623, considering that V~sin~i$\sim$ 70kms$^{-1}$, i=60$\pm$10$^o$, and 160 kms$^{-1}$$<$ v$_c$$<$ 250 kms$^{-1}$, the rotational velocity v$_{rot}$ ranged between 0.3 and 0.6 v$_c$, which is not large enough to reach the bi-stability limit and produce a dense, optically thick equatorial region. \textbf{Moreover, the efficiency of the bi-stability model also increases with the luminosity as shown by Pelupessy et al. (2000). For instance, for a 20000~K star rotating at 60$\%$ of its critical velocity, the polar-to-equatorial density ratio is 0.95 for L$_\star$~=~10$^{4.5}$~L$_\odot$ while it decrease down to 0.21 for L$_\star$~=~10$^{6.0}$~L$_\odot$ .} Finally, other phenomena have to be present to explain the differences between a standard A-type supergiant and HD~62623.

\subsection{The putative binary nature of HD~62623}

Studing the periodical variations of HD~62623's radial velocity (137.7 or 161.1 days), Plets et al. (1995) found evidence of the binarity of this star. They concluded that the system is composed of a massive primary; i.e.,  M$_1$=31M$_\odot$ or M$_1$=39M$_\odot$, and a small mass ratio of M$_2$/M$_1$=0.03 or M$_2$/M$_1$=0.15. The estimation of the projected semi-major axis of the system is 1.6~AU~$<$~a.sin~i~$<$~2.4~AU. Considering that the companion orbits in the equatorial plan of HD~62623's primary, and using our estimation for the inclination angle i~=60~$\pm$~10$^o$, we obtain a semi-major axis of a~=~2.3$~\pm$~0.6~AU.

Finally, Plets et al. (1995) proposed a general scheme for HD~62623.  The matter is probably ejected by the evolved massive primary stellar wind with typical terminal velocity of 80~kms$^{-1}$. However, the matter located in the equatorial plane cannot reach this value since it is decelerated by the companion, finally forming a circumbinary disk. Thanks to this deceleration, the density in the equatorial plane could be 10 times higher than the typical density for a single A2I star stellar wind.

Binarity can be detected using interferometric techniques, and recent VLTI/AMBER observations of the classical Be star $\delta$~Cen by Meilland et al. (2008) and VLTI/AMBER + VLTI/MIDI observations of the unclassified B[e] HD~87643 by Millour et al. (2009) resulted in the discovery of companions around these two stars. However, HD~62623's companion cannot be detected in our VLTI/MIDI data since the N band flux is fully dominated by the dusty disk emission and because the projected separation between the two components of the system is too small compared to the spatial frequencies probed by the instrument.

Nevertheless, considering the stellar parameters determined by Plets et al. (1995), the luminosity of the central star is at least 14 times larger than the companion's, since for M$_1$=31M$_{\odot}$ and M$_2$/M$_1$=0.15, we obtain M$_2$=5M$_\odot$, and the companion spectral class is between B8III and B6V. Thus, it would result in visibility oscillations with a maximum amplitude of 2/14 $\sim$ 0.14 for the binary system assuming no envelope contribution. In the H and K band, the envelope emission should represent 20$\%$ and 60$\%$ of the total flux, respectively. Thus, the oscillations should have a maximum amplitude of 0.11 in the H band and 0.056 in the K band and could be detected, at least in the H band, with the VLTI/AMBER.

Finally, it is interesting to note that the putative companion is probably located inside the inner radius of the dusty circumstellar disk, as seen by the VLTI/MIDI instrument. The disk geometry determined using the MC3D code and our interferometric observations is thus compatible with Plets et al.'s (1995) scheme for HD~62623.

\section{Conclusion}

Thanks to these first VLTI/MIDI observations of HD~62623, we were not only able to determine the extension and flattening of a supergiant star showing the B[e] phenomenon, as was already done by Domiciano de Souza et al. (2007) on CPD -57$^o$ 2874, but also to significantly constrain the inner rim of the dusty disk and the disk opening angle. However, we were not able to infer the possible disk flaring.

We have shown that the inner gaseous envelope likely acts as a reprocessing disk. Since we detect residual visibilities for the 70 m baseline, it is also probable that the inner gaseous envelope free-free N-band emission is not negligible and contributes to 10-20$\%$ of the total flux. However, other hypothesis like clumping within the disk cannot be totally ruled out.

How dust can form around this highly luminous (i.e.,~L$_\star\sim$~17000~L$_\odot$) star still remains unclear. Moreover, HD~62623 is one of the few known A[e] supergiants, and it is hard to understand what the differences are between this object and other stars with similar spectral classes, like Deneb, which only exhibits a moderate radiatively driven stellar wind and no dust in its circumstellar environment. Lamers \& Pauldrach's (1997) bi-stability model does not seem to fit accurately HD~62623 data since this star is likely to have a moderate rotational velocity, i.e. 0.3$<$V$_c$$<$0.6. 

Moreover, comparing the dusty disk inner rim extension R$_{in}$~=~3.8$~\pm$~0.7~AU and Plets et al.'s (1995) companion orbit estimation; i.e., 1.6~AU~$<$~a.sin~i~$<$~2.4~AU, we can conclude that a stellar wind deceleration by the companion's gravitational effects remains the most favorable hypothesis.

However, to definitely rule out other possible explanations, we need to constrain both the geometry and kinematics of the inner gaseous envelope. As already shown for classical Be stars (Meilland et al. 2007a, 2007b) and Herbig stars (Malbet et al. 2007 ; Kraus et al. 2008), spectrally resolved interferometry with the VLTI/AMBER instrument in medium (R=1500) or high (R=12000) spectral resolution mode is the most suitable method for that purpose. New VLTI/MIDI observations with baselines longer than 70~m can also bring additional constraints on the gaseous environment geometry. Finally, as shown for Be stars by Meilland et al. (2007a ; 2009) and Kervella et al. (2009), comparing N band VLTI/MIDI and H-K-band VLTI/AMBER measurements can add additional constraints on the physical parameters, for instance the temperature and density of the gaseous environment.

\begin{acknowledgements}
The Programme National de Physique Stellaire (PNPS) and the Institut National en Sciences de l'Univers (INSU) are acknowledged for their financial supports. F. Millour \& A. Meilland acknowledge financial support from the Max Planck Institut f\"ur Radioastronomy.
\end{acknowledgements}

\end{document}